\documentclass[fleqn,usenatbib]{mnras}

\usepackage{newtxtext,newtxmath}

\usepackage[T1]{fontenc}

\DeclareRobustCommand{\VAN}[3]{#2}
\let\VANthebibliography\thebibliography
\def\thebibliography{\DeclareRobustCommand{\VAN}[3]{##3}\VANthebibliography}

\usepackage{graphicx}	
\usepackage{amsmath}	
\usepackage{url}
\usepackage[dvipsnames]{xcolor}

\newcommand{\review}[1]{{#1}}
\newcommand{\reviewb}[1]{{#1}}
\newcommand{\reviewc}[1]{{#1}}

\title[FRB 20121102A Survey]{A broad survey of spectro-temporal properties from FRB 20121102A}

\author[M. A. Chamma et al.]{
Mohammed A. Chamma,$^{1}$\thanks{E-mail: mchamma@uwo.ca}
Fereshteh Rajabi,$^{1,2}$
Aishwarya Kumar$^{1}$
and Martin Houde$^{1}$\thanks{E-mail: mhoude2@uwo.ca}
\\
$^{1}$Department of Physics and Astronomy, The University of Western Ontario, 1151 Richmond Street, London, Ontario N6A 3K7, Canada\\
$^{2}$Perimeter Institute for Theoretical Physics, Waterloo, Ontario N2L 2Y5, Canada
}

\pubyear{2023}

\begin{document}
\label{firstpage}
\pagerange{\pageref{firstpage}--\pageref{lastpage}}
\maketitle

\begin{abstract}
We survey the spectro-temporal properties of fast radio bursts from FRB~20121102A \review{observed by earlier studies} across a wide range of frequencies. We investigate 167 bursts from FRB~20121102A spanning frequencies 1--7.5GHz, durations of less than 1 ms to approximately 10 ms, with low and high energies, and with \review{different} wait-times. \review{We find from this sample of bursts} a strong agreement with the inverse relationship between sub-burst slope and duration and with other predictions made by the triggered relativistic dynamical model (TRDM). Earlier results found agreement with those predictions across three different repeating FRB sources. For this sample of bursts, we find that the sub-burst slope as well as the `sad trombone' drift rate are consistent with being \review{in a} quadratic \review{relationship} with frequency \review{and that} both these quantities are inversely proportional to the duration. \review{We also find that} the duration decreases with increasing frequency as well as a statistically significant correlation between the sub-burst duration and bandwidth (proportional to $t^{-1/2}$) that is unexpected. No distinct group of bursts in this sample deviated from these relationships, however significant \review{scatter} can be seen in measurements. This study demonstrates the consistent existence of relationships between the spectro-temporal properties of bursts from a \review{repeating} FRB source\review{. A} simple explanation for the inverse relation between the sub-burst slope and duration is an inherently narrowband emission process. We make \review{all measurements} available as well as a graphical user interface called \textsc{Frbgui} developed and used to perform measurements of burst waterfalls.
\end{abstract}

\begin{keywords}
(transients:) fast radio bursts -- methods: data analysis -- relativistic processes -- radiation: dynamics -- radiation mechanisms: non-thermal
\end{keywords}



\section{Introduction}
Fast radio bursts (FRBs) are short and intense pulses of radiation whose emission mechanism eludes understanding despite a multitude of theoretical models and many recent observational efforts. Observationally, the originating environments, energy distributions, polarization properties, and activity cycles of FRB sources have been investigated\review{,} with each new study often adding a new unexpected characteristic \citep[\reviewb{see}][\reviewb{for a recent review}]{Petroff2022}. Theoretical explanations for FRBs center around extreme environments such as magnetars where large numbers of particles in a plasma state emit coherently \citep{Platts2019,Lyubarsky2021}. \reviewb{In recent years, hundreds of new sources have been discovered with a clear distinction in properties between repeating and non-repeating sources \citep{CHIMEFRB2021,Pleunis2021a}.} This flurry of activity around FRBs has revealed significant challenges in understanding this phenomena and in constraining the possible explanations.


Many of the observational characteristics of FRBs vary dramatically from source to source, making it unclear which features are arising due to the emission mechanism, the environment, or propagation effects. The spectral luminosities of FRB sources span several orders of magnitude and the durations of bursts can range from tens of nanoseconds to tens of milliseconds \citep{Nimmo2022}. The polarization properties of FRBs vary as well; for example in FRB~20121102A \citep{Spitler2014,Spitler2016} bursts have a constant polarization across their duration \citep{Michilli2018}, whereas in FRB~20180301A the polarization angle of some bursts show diverse behaviours \citep{Luo2020}. In FRB~20190520B, the \reviewb{observed} rotation measures (RMs) \reviewb{exhibited changes in sign} \review{\citep{AnnaThomas2022}, while FRB~20121102A has the largest absolute RM \review{and range of RMs} observed \citep{Hilmarsson2021}}. These characteristics of FRB sources complicate our understanding \review{of the processes involved} and are likely due to an overlap of multiple different phenomena.

One avenue for understanding the emission mechanism of FRBs is to study the spectro-temporal properties of bursts, which has revealed several relationships that are common from burst to burst and even from source to source. Among these quantities, which include the bandwidth, duration, and central frequency, are the drift rate and the similar but distinct sub-burst slope or intra-burst drift\footnote{The `sub-burst slope' terminology was used in \citet{Chamma2021} while `intra-burst drift' was used in \citet{Jahns2022}. Both terms describe the same measurement and we will use `sub-burst slope' hereafter.}. The drift rate refers to the change in frequency of multiple resolved sub-bursts within a single waterfall, and the tendency for later sub-bursts to arrive at lower frequencies is called the `sad trombone' effect. The sub-burst slope on the other hand refers to the change in frequency with time within a single sub-burst or pulse. \citet{Hessels2019} studied bursts from FRB~20121102A and the relationship between their frequency and the drift rate of multiple resolved sub-bursts, finding that the drift rate increased with frequency. This relationship appeared to be linear \citep{Josephy2019}. 

\reviewc{Modelling and understanding the origins of spectro-temporal relationships is an informative avenue for understanding FRBs.} In the triggered relativistic dynamical model (TRDM) proposed by \citet{Rajabi2020}, an inverse relationship was predicted between a sub-burst's slope and its duration from the assumptions of a narrow-band emission process and \reviewc{up-to-}relativistic motions in the source. This prediction agreed with measurements made for a sample of bursts from FRB~20121102A. This relationship was further explored in \citet{Chamma2021} where bursts analysed from three repeater sources (FRB~20180916B \citep{CHIME2020b} and FRB~20180814A \citep{Amiri2019} in addition to FRB~20121102A) also had sub-burst slopes that varied inversely with duration, indicating that the same relationship could describe the features of bursts from different sources. The model in \citet{Rajabi2020} also predicted a quadratic relationship between the sub-burst slope and the frequency, and some evidence of this can be seen in various datasets as plotted in Fig. 5 of \citet{Chamma2021} and Fig. 8 of \citet{Wang2022}. \citet{Jahns2022} studied over 800 bursts from FRB~20121102A in the 1.1-1.7 GHz band and also found the sub-burst slope to be inversely proportional to duration. In addition they compared the sub-burst slope to a dozen drift rates and found that the drift rates were larger but seemed to extend the same trend with the duration as the sub-burst slopes do \citep{Jahns2022}. This behaviour for drift rates and sub-burst slopes obeying similar or identical relationships is expected within the TRDM when groups of sub-bursts are emitted at roughly the same time \citep[Sec. 3.1 of][]{Chamma2021, Rajabi2020}. Given these recent discoveries it is fruitful to study the spectro-temporal features of bursts in order to better characterize these relationships, to find their limitations, and to find possible commonalities between sources that will contribute to our understanding of the FRB emission mechanism.

\reviewc{Studies thus far have focused on samples of bursts that are observed in relatively narrow frequency ranges (limited by the observing instrument), and often spectro-temporal measurements are not performed. While studies that perform such measurements exist using deep samples of bursts (e.g. \citealt{Aggarwal2021,Pleunis2021a,Jahns2022}), there is a dearth of studies that broadly and comprehensively sample the properties of all of a source's bursts. This leaves a potential gap in our knowledge of the spectro-temporal relationships that exist and their applicability to a broad sample of a source's bursts.}


\reviewc{In this work we collect a large and diverse sample of bursts from a single source and search for spectro-temporal relationships, while also investigating if the same relationships, such as that between the sub-burst slope and the duration, are verified across the entire sample.} \review{T}his will indicate if bursts from a single source can be differentiated based on the spectro-temporal relationships they obey as well as the robustness of these relationships. To this end we collect bursts from multiple observational studies of the repeating source FRB~20121102A and measure their spectro-temporal properties. As one of the best observed repeaters, FRB~20121102A \reviewc{has been localized to a star-forming dwarf galaxy with a persistent radio counterpart and has bursts that cover a wide range of frequencies and durations \citep{Bassa2017,Tendulkar2017,Chatterjee2017,Marcote2017}}. Our measurements include the central frequency, the bandwidth, the sub-burst slope, the sad-trombone drift rate (when applicable), and the duration of every burst and sub-burst\review{. W}e investigate the relationships between these quantities, \review{and} \reviewc{whether} a single relationship is sufficient to describe the data or if deviations exist. \reviewc{Understanding the scope of these relationships and how well they describe the complete cohort of bursts from a source can provide valuable information about the emission mechanism and host environment.}

In the following, Section \ref{sec:sampled} will describe the observations used and the bursts sampled for this study. Section \ref{sec:methods} will describe how data are prepared, how measurements are obtained using 2D autocorrelations, and how the dispersion measure (DM) for each burst is handled and optimized within the context of the burst sample. Section \ref{sec:results} will describe the measurements obtained and explore different correlations between the spectro-temporal properties of the bursts while Section \ref{sec:discussion} will discuss the relationships observed, the predictions of the TRDM in more detail, and possible \reviewc{interpretations and} implications. The paper is summarized with our conclusions in Section \ref{sec:conclusions}.


\section{Burst Sampling}\label{sec:sampled}

We \reviewc{describe} here the observations used and the properties of the bursts sampled for our study.

\cite{Michilli2018} observed 16 bursts from FRB~20121102A using the Arecibo Observatory in a band that spanned 4.1--4.9 GHz. We use all the bursts they observed and separate the components of three of their bursts (M9, M10, and M13) for a total of 19 single pulses. The durations of these bursts \reviewc{range from 0.03--1.36 ms}, as measured by their full-width-half-maximum (FWHM) and excluding the bursts whose components we have separated \citep{Michilli2018}.

\cite{Gajjar2018} observed 21 bursts from FRB~20121102A using the Green Bank Telescope in a band of 4--8 GHz all of which were 100\% linearly polarized. Of these observations we exclude 5 due to their low SNR \reviewc{(which prevents or complicates measurements)} and split three of their bursts (11A, 12A, and 12B) for a total of 21 single pulses, with durations that range \review{from} 0.18--1.74 ms \citep{Gajjar2018}.

\cite{Oostrum2020} observed 30 bursts from FRB~20121102A using the WSRT/Apertif telescope in a band spanning 1250--1450 MHz (and 1220--1520 MHz for one burst) with much lower levels of linear polarization than observed at higher frequencies. We use 23 of these bursts with the remaining seven excluded due to low SNR\review{. A}ll bursts from this dataset were single pulses. The durations of these bursts, as measured by a top hat pulse with an equivalent integrated flux density, span 1.6--8.2 ms \citep{Oostrum2020}. 

\cite{Aggarwal2021} searched data \review{observed} by \cite{Gourdji2019} using the Arecibo telescope in a band spanning \review{580} MHz (\review{with a backend spanning 800 MHz}) and centered around 1375 MHz\review{. I}ncluding the 41 bursts found by \cite{Gourdji2019}, \review{the search yielded} a total of 133 bursts in three hours of data. Notably, almost all of these bursts occur above 1300 MHz. Of these bursts we exclude almost half due to a low SNR and separate 6 into single pulses for a total of 63 bursts. The durations of the bursts used span 1.16--17.16 ms based on the FWHM obtained from the burst autocorrelation (see Section \ref{sec:methods}).


\cite{Li2021} used the FAST telescope to detect 1652 bursts in about 60 hours of data over 47 days in a band spanning 1000--1500 MHz. These bursts followed a bimodal energy distribution\footnote{\review{For a discussion on differences in methodology when using the observing bandwidth versus burst bandwidth to compute burst energies see \cite{Aggarwal2021b} and Section 3 of \cite{Jahns2022}.}} with peaks around $10^{37.8}$ and $10^{38.6}$ erg as well as a bimodal wait-time distribution with peaks at around 3.4 ms and 70 s. \citet{Jahns2022} found a weak bimodality in the burst energy distribution in their sample of 849 bursts and could not \review{conclusively} confirm the result of \cite{Li2021}. 

In order to investigate if these differing properties \review{of wait-time and energy might} correspond to different spectro-temporal features \review{or relationships we sampled bursts from both peaks of the bimodal energy and wait-time distributions reported by \cite{Li2021}. For the energy distribution} we sampled 20 bursts from both peaks (10 each) of the energy distribution by filtering their list of bursts to those with estimated energies between $10^{37.7}$ and $10^{37.8}$ erg for the first peak and $10^{38.6}$ and $10^{38.7}$ erg for the second peak. To select bursts with high enough SNR for a good measurement we additionally filter bursts from the first peak that have a peak flux density greater than 10 mJy and greater than 100 mJy from the second peak. \review{Different flux limits are chosen for each peak to ensure a large enough sample and are somewhat arbitrary.} \review{For} the wait-time distribution, \review{we filtered} for bursts with wait-times between $\sim$4 to $\sim$6 ms and with a peak flux density above 40 mJy, and wait-times between $\sim$63 to $\sim$100 s above 100 mJy. This yielded 11 and 13 bursts from each peak, respectively, for an additional 24 bursts. Of this sample, several more were excluded due to an SNR that was still too low to obtain measurements and many of the short wait-time bursts were split into multiple components, which finally resulted in a total of 41 single pulses. \review{Of this total, 6 and 12 bursts were from the short and long wait-time peaks, respectively. Only 2 bursts remained from the low energy peak and 7 from the high energy peak, 1 burst was included for having a long duration, and 13 were new pulses that we had separated. Due to challenges in acquiring the data we decided to proceed with this sample.} The bursts used span a frequency range of 1080--1430 MHz and their FWHM durations span 0.56--15.43 ms.

\begin{figure}
    \centering
    \includegraphics[width=\columnwidth]{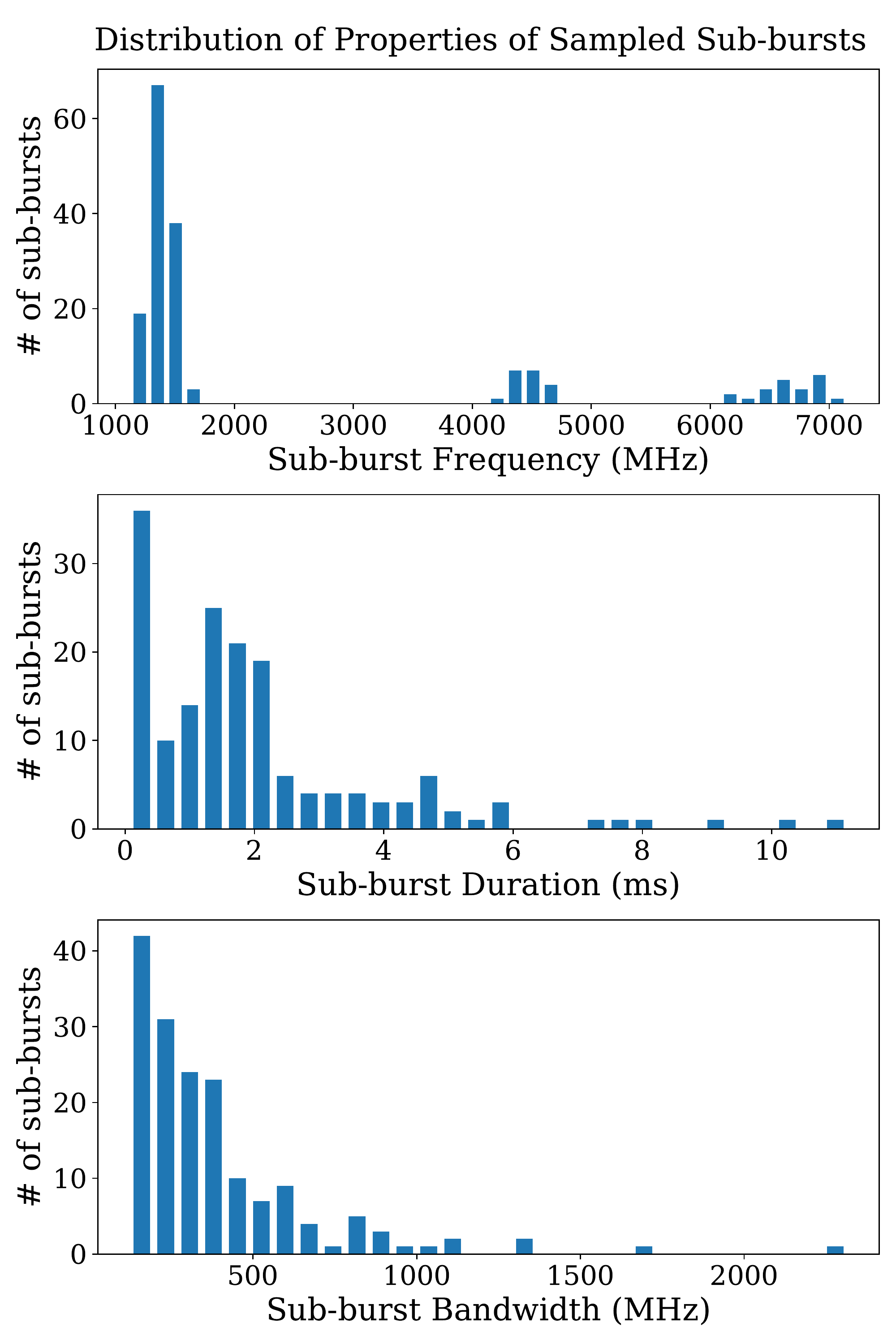}
    \caption{Histograms of the sub-bursts sampled for this study showing the frequency (top), sub-burst duration (middle), and sub-burst bandwidth (bottom).}
    \label{fig:sampledist}
\end{figure}

The sample of bursts analysed in this study total 167 and broadly represent all the types of bursts that have been observed from FRB~20121102A, spanning frequencies ranging from 1080 MHz to 7.4 GHz and durations from less than 1 ms to about and greater than 10 ms. Figure \ref{fig:sampledist} shows a distribution of the frequencies and durations of the bursts used in our sample. 

\section{Methods and Analysis}\label{sec:methods}

This section will describe how the burst waterfalls are loaded and their spectro-temporal properties measured via 2D autocorrelations of the waterfall \citep{Hessels2019,Chamma2021}, as well as the dispersion measure (DM) ranges used for the measurements, and how measurements are reviewed and validated. We also describe a graphical user interface (GUI) we developed to aid in the measurement of burst properties that is easily extensible and publicly available.

The FRB observations described in Section \ref{sec:sampled} were obtained from the authors of their respective publications and we \review{describe in the following how they are} pre-process\review{ed} in order to \review{reduce computation time}, remove radio frequency interference (RFI), and/or crop burst \review{files} before performing measurements. FRB waterfall data are available in various formats and include \texttt{filterbank} and \texttt{PSRFITS} files, which were loaded using the \textsc{PyPulse}\footnote{\url{https://github.com/mtlam/PyPulse}} and \textsc{Your}\footnote{\url{https://github.com/thepetabyteproject/your}} software packages before being stored as 2D Python \texttt{numpy} arrays. 

The waterfalls of bursts from \cite{Michilli2018} were provided via an ASCII dump, dedispersed, and at frequency and time resolutions of 1.5625 MHz and 0.01024 ms, respectively, with 512 frequency channels and 512 time samples (i.e. with dimensions 512 $\times$ 512). 

Burst waterfalls from \cite{Gajjar2018} were available dedispersed in \texttt{PSRFITS} format at a resolution of 183 kHz and 0.01024 ms with 19456 frequency channels and 2048 time samples, which we subsampled in frequency by a factor of 8 to obtain 2432 frequency channels at a resolution of 1.464 MHz. These were then stored in arrays of size 1690 $\times$ 2048 to exclude masked channels present at the bottom of the band. 

Waterfalls from \cite{Oostrum2020} were obtained un-dedispersed in the \texttt{PSRFITS} format with resolutions of about 0.195 MHz and 0.04096 ms with size 1024 $\times$ 25000 channels, which we dedispersed to each burst's reported DM and downsampled by a factor of 8 to resolutions of 1.5625 MHz and 0.32768 ms. These waterfalls were then centered and cropped to be stored at a size of 128 $\times$ 200 channels. In addition, before downsampling we applied both a spectral kurtosis and Savitzky-Golay filter \citep[SK-SG filter;][]{Agarwal2020, Nita2016} that is available via the \textsc{Your} package to mask channels with high RFI. 

The waterfall data from \cite{Aggarwal2021} are provided in long duration filterbank files with a list of candidate timestamps, and the \textsc{Burstfit}\footnote{\url{https://thepetabyteproject.github.io/burstfit/}} package is used to dedisperse the data to each burst's DM and select a 0.1 s long waterfall around the burst. These waterfalls are then saved as arrays of size 64 $\times$ 1220 at resolutions of 12.5 MHz and 0.08192 ms, respectively. 

Data from \cite{Li2021} were obtained via correspondence and consisted of \texttt{PSRFITS} files for each burst. We load these with \textsc{Your} at a size of 4096 $\times$ 131072 channels and resolutions of 122 kHz and 0.098304 ms, and then filter them with an SK-SG filter. We then dedisperse each waterfwall to the reported burst DM, subsample to 256 frequency channels, center about the time channel with the peak frequency-averaged intensity and crop to an array of size of 256 $\times$ 1000 with resolutions of 1.952 MHz and 0.098304 ms. 

All the bursts in this study are preprocessed in the way described in order to facilitate the measurements of their spectro-temporal features.

Despite the pre-processing of the waterfalls there are still several tasks that are needed on a burst-by-burst basis, that, with a large number of bursts and measurements to manage, can quickly become overwhelming and difficult to review. These tasks include additional noise removal \review{(such as masking channels or applying an SK-SG filter)}, additional subsampling to increase the S/N, and separating the components of bursts with multiple pulses. While these tasks can be automated to an extent, being able to manage and \reviewc{refine} a measurement allows for more accurate results and the inclusion of strange bursts that might not fit in an automation pipeline of limited complexity. To this end we developed an extensible graphical user interface (GUI) called \textsc{Frbgui} that allows a user to input additional masks, change the subsampling of the data, and isolate components of a burst, and used it to prepare bursts and obtain measurements of the spectro-temporal features.

\begin{figure*}
    \centering
    \includegraphics[width=0.9\textwidth]{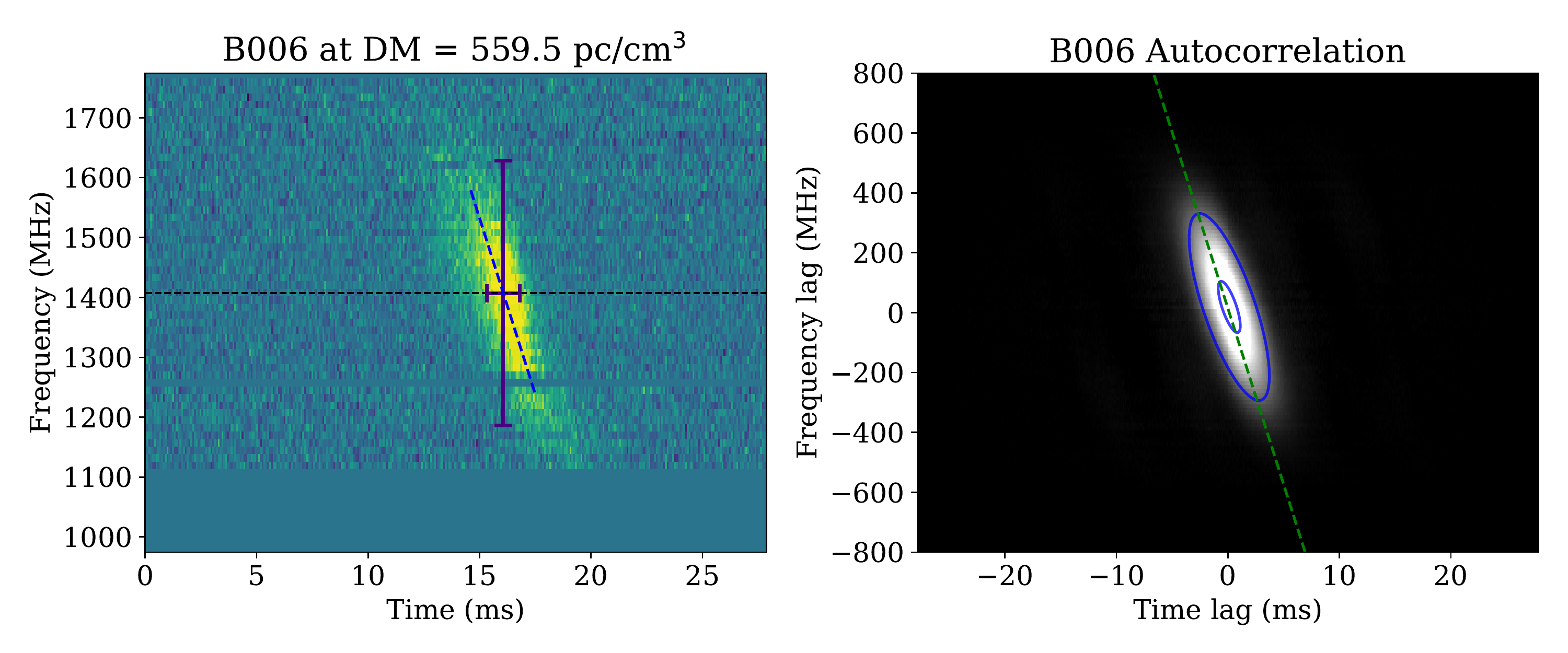}
    \caption{Example measurement of burst B006 from \protect\cite{Aggarwal2021}. The left panel shows the burst waterfall dedispersed to 559.5 pc/cm$^3$ with the dashed line, horizontal bar, and vertical bar indicating the sub-burst slope, duration, and \reviewc{burst} bandwidth obtained via the 2D Gaussian model of the autocorrelation, shown on the right. On the right, the image shows the computed autocorrelation, and the blue outlines are contours of the Gaussian model at a quarter and 90\% of the peak. The dashed green line shows the corresponding slope.}
    \label{fig:B006}
\end{figure*}

The spectro-temporal features of each burst are obtained via a 2D Gaussian fit to the 2D autocorrelation of the burst waterfall. An autocorrelation of the waterfall helps increase the S/N for measurement and limits the effect of spectral structures, noise, and banding in the burst. The Gaussian model of the autocorrelation therefore provides an analytical and robust way of measuring the spectro-temporal features from a small number of parameters. This technique is detailed in, \reviewb{for example,} Appendix A of \cite{Chamma2021} (\reviewb{see \citealt{Jahns2022} for alternative formulations}), and for this study, the Gaussian fit is obtained using the normalized physical coordinates of the autocorrelation. Thus, the dimensions of the Gaussian model's parameters are unitless, and we write with $x=t/(1\,\text{ms})$ and $y=\nu / (1\,\text{MHz})$
\begin{align}
    G\left(x,y\right) & = C\exp\left\{-\frac{1}{2}\left[(x-x_0)^2\left(\frac{\cos^{2}\theta}{b^2}+\frac{\sin^{2}\theta}{a^2}\right)\right.\right.\nonumber\\
    & \rule{0mm}{6mm}+2(x-x_0)(y-y_0)\sin\theta\cos\theta\left(\frac{1}{b^2}-\frac{1}{a^2}\right)\nonumber\\
    & \left.\rule{0mm}{6mm}\left.+(y-y_0)^2\left(\frac{\sin^{2}\theta}{b^2}+\frac{\cos^{2}\theta}{a^2}\right)\right]\right\} ,\label{eq:Gaussian}
\end{align}
where $C$, $x_0$, $y_0$, $a$, $b$, and $\theta$ are the model parameters corresponding to the amplitude, central $x-$ and $y-$ positions, the standard deviations of the Gaussian, and the orientation of the semi-major axis ($a$) measured counterclockwise from the positive $y-$axis. The fit is found \review{with} the \texttt{scipy.optimize.curve\_fit} package \review{using the normalized physical coordinates defined above}. The sub-burst slope and duration are obtained via equations A2-A3 of \cite{Chamma2021}, with the modification that the unit conversion becomes unity due to the choice of coordinates\footnote{\reviewc{When using coordinates numbered by channel fits did not always converge or were inaccurate. Using normalized physical coordinates resolved this issue.}} when obtaining the Gaussian fit, so that
\begin{align}
    \frac{d\nu_{\text{obs}}}{dt_{\text{D}}} &= -\left(1\,\frac{\text{MHz}}{\text{ms}}\right)\,\cot\theta,\label{eq:slope}\\
    t_\text{w} &= \left(1\,\text{ms}\right)\,\frac{ab}{\sqrt{b^{2}\sin^{2}\theta+a^{2}\cos^{2}\theta}},\label{eq:tw}
\end{align}
where $d\nu_{\text{obs}}/dt_\text{D}$ and $t_\text{w}$ are the sub-burst slope and sub-burst duration\footnote{The duration defined in eq. (\ref{eq:tw}) is the correlation length of the burst, and can be converted to other definitions of burst duration with a simple scaling. If the burst is Gaussian with a standard deviation of $\sigma_\text{p}$ and FWHM $t_{\text{FWHM}} = 2\sqrt{2\ln{2}}\sigma_\text{p}$, then the correlation length $t_\text{w}$ is related to those durations by $t_\text{w} = \sqrt{2}\,\sigma_\text{p}$, and $t_\text{w} = 1/\sqrt{4\ln{2}}\,t_\text{FWHM} \simeq 0.6\, t_\text{FWHM}$.}, respectively, and $\nu_\text{obs}$ and $t_\text{D}$ are the observed frequency and delay time (or arrival time) of the burst, written in the formalism of \cite{Rajabi2020}. We also compute the total bandwidth $B_\text{tot}$ according to
\begin{align}
    B_{\text{tot}}=(1\,\text{MHz})\,\sqrt{8\ln2}\,a\cos\theta,\label{eq:bandwidth}
\end{align}
which is the semi-major axis of the Gaussian ellipsoid scaled to its FWHM value and projected on to the frequency axis. Figure \ref{fig:B006} shows an example measurement of the spectro-temporal properties of burst B006 from \cite{Aggarwal2021}. For bursts with multiple components in a single waterfall, the autocorrelation changes in size to include the multiple components and equations (\ref{eq:slope}) to (\ref{eq:bandwidth}) are still valid. We can therefore perform the same analysis to obtain the drift rate if a fit can be found to the larger autocorrelation. The drift rates we obtain are treated distinctly from the sub-burst slopes as these potentially arise from different phenomena. 

Because the choice of DM affects the value of the sub-burst slope and, to a lesser extent, the sub-burst duration, each burst is measured over a range of trial DMs. We measure each burst over a grid of DMs spanning 555 to 575 pc/cm$^3$ in steps of 0.5 pc/cm$^3$, chosen to account for the historical range of DMs observed from FRB~20121102A. Each burst is incoherently dedispersed from its reported DM to the trial DM and measured via autocorrelation, resulting in 42 sets of measurements per burst including the reported burst DM. 

The total set of measurements obtained over the DM grid are then filtered to exclude sub-burst slopes that are positive or with \review{measurement uncertainties} larger than 40\%. We exclude all positive sub-burst slopes under the assumption that they are unphysical and due to over dedispersion \citep{Chamma2021, Jahns2022}. Most of the \review{measurement uncertainties} larger than 40\% are due to the maximal DM in the range being greater than the \review{published} DM of a burst. Each burst therefore has a sub-burst slope that is nearly vertical at some point over the DM range where \review{uncertainties} can be very large (see eq. \ref{eq:slope}).

\reviewb{For each dataset, t}he remaining measurements are then grouped by DM and a fit is found between the sub-burst slope and duration with the form $(d\nu_{\text{obs}}/dt_{\text{D}}) / \nu_{\text{obs}} = A/t_\text{w}$, which is the relationship predicted for these two properties by the TRDM. Of this set of fits, we compute a reduced-$\chi^2$ to assess the goodness of the fit and tabulate the remaining number of bursts \review{that passed the filtering process} for each DM, \reviewb{by dataset}. For each of the datasets listed in Section \ref{sec:sampled}, the range of DMs is \reviewc{then} limited to only include those DMs that include all the bursts in the sample, and, within this limited range, the DM with optimal (\reviewc{closest to unity}) reduced-$\chi^2$ is chosen as the representative DM for that sample. 

The range of values in the remaining measurements in the limited DM range are then used as an estimate of the uncertainties. For example, the bursts from \cite{Michilli2018} are measured over the DM range 555--575 pc/cm$^3$, and, after filtering invalid measurements, the remaining DMs that include all of the bursts range from 555--560 pc/cm$^3$. Within this limited DM range, the DM with optimal reduced-$\chi^2$ is found to be 558 pc/cm$^3$, and the range of measurements over the limited DM range are used as the uncertainties of the measurement at 558 pc/cm$^3$. Treating the measurements of each burst in this manner helps account for the effect of the DM on the measurement and allows for an estimate of the uncertainties beyond those from the Gaussian model. \review{To summarize, this measurement filtering process described in the preceding paragraphs includes the exclusion of positive sub-burst slopes and slopes with large uncertainties, the determination of the maximum valid and optimal DMs using the sub-burst slope law, and the determination of the \reviewc{range of} measurement \reviewc{values} over the valid DM range.}

The measurements are reviewed visually with the help of the bars and slope indicators such as those shown on both panels of Figure \ref{fig:B006}. Plots and tables of all measurements can be found online at \href{https://github.com/mef51/SurveyFRB20121102A}{github.com/mef51/SurveyFRB20121102A}.







\section{Results}\label{sec:results}
We describe the results of the autocorrelation analysis, the measurement filtering process, and the relationships between the spectro-temporal properties in this section. The sub-burst slope (normalized by the observing frequency) vs. sub-burst duration is shown in Figure \ref{fig:SlopevDuration}. Figures \ref{fig:PropertiesvFreq} to \ref{fig:PropertiesvBand} show different (potential) correlations between the measured spectro-temporal properties, including the sub-burst duration and sub-burst slope with frequency, and correlations between the bandwidth with frequency, duration and sub-burst slope. Figure \ref{fig:Drifts} examines the small sample of bursts with multiple components and shows the drift rate vs. duration and drift rate vs. frequency. In Figure \ref{fig:PropertiesvBand} we find an unexpected correlation between the bandwidth and duration (and thus sub-burst slope). Finally we look in particular at \review{the spectro-temporal features of} bursts from \cite{Li2021} in Figure \ref{fig:Li} and the \review{absence of correlations} between bursts from different energy and wait-time peaks.

As mentioned in the previous section, the measurements are filtered to exclude invalid values and this process results in some bursts not having a valid measurement at a particular DM, which results in a limited DM range after requiring that all bursts are included. After this filtering, the ranges of DMs for which bursts have valid measurement varied from dataset to dataset; while all five datasets had valid measurements down to 555 pc/cm$^{3}$, the highest DMs at which all bursts in a dataset  still had valid measurements differed. These maximal DMs along with the DM in that range that had the optimal fit between the sub-burst slope and duration are listed in Table \ref{tab:dmresults}. The average maximal DM (weighted by the number of bursts in each dataset) was 561.5(4) pc/cm$^{3}$ and the average optimal DM was 560.1(8) pc/cm$^{3}$. In the figures to follow, burst measurements are always displayed at the optimal DM of the dataset they come from with uncertainties estimated from the range of measurements found within the limited DM range.
\begin{table}
    \begin{centering}
    \begin{tabular}{rlll}
    \hline
    \textbf{Dataset} & \textbf{Max valid DM} & \textbf{Optimal DM} & \textbf{\# of bursts}\\& \textbf{(pc/cm$^{3}$)} & \textbf{(pc/cm$^{3}$)}\tabularnewline
    \hline
    \hline
    \cite{Michilli2018} & 560.0 & 558.0 & 19\tabularnewline
    \hline
    \cite{Gajjar2018} & 563.5 & 557.5 & 21\tabularnewline
    \hline
    \cite{Oostrum2020} & 563.0 & 563.0 & 23\tabularnewline
    \hline
    \cite{Aggarwal2021} & 560.5 & 559.5 & 63\tabularnewline
    \hline
    \cite{Li2021} & 562.0 & 562.0 & 41\tabularnewline
    \hline
    \textbf{Weighted Average:} & 561.5(4) & 560.1(8) & 167\tabularnewline
    \hline
    \end{tabular}
    \par\end{centering}
    \caption{Results from the measurement exclusion and DM optimization process
    described in Section 3. The maximum valid DM is the DM beyond which bursts
    start being excluded for having invalid sub-burst slope measurements
    (ie. being over-dedispersed). The optimal DM is the DM for which the
    reduced-$\chi^{2}$ of the fit between the sub-burst slope and duration
    measurements at that DM is closest to unity. The DM step size used was 0.5 pc/cm$^3$. }\label{tab:dmresults}
\end{table}

\subsection{Spectro-temporal Properties}
\begin{figure*}
    \centering
    \includegraphics[width=\textwidth]{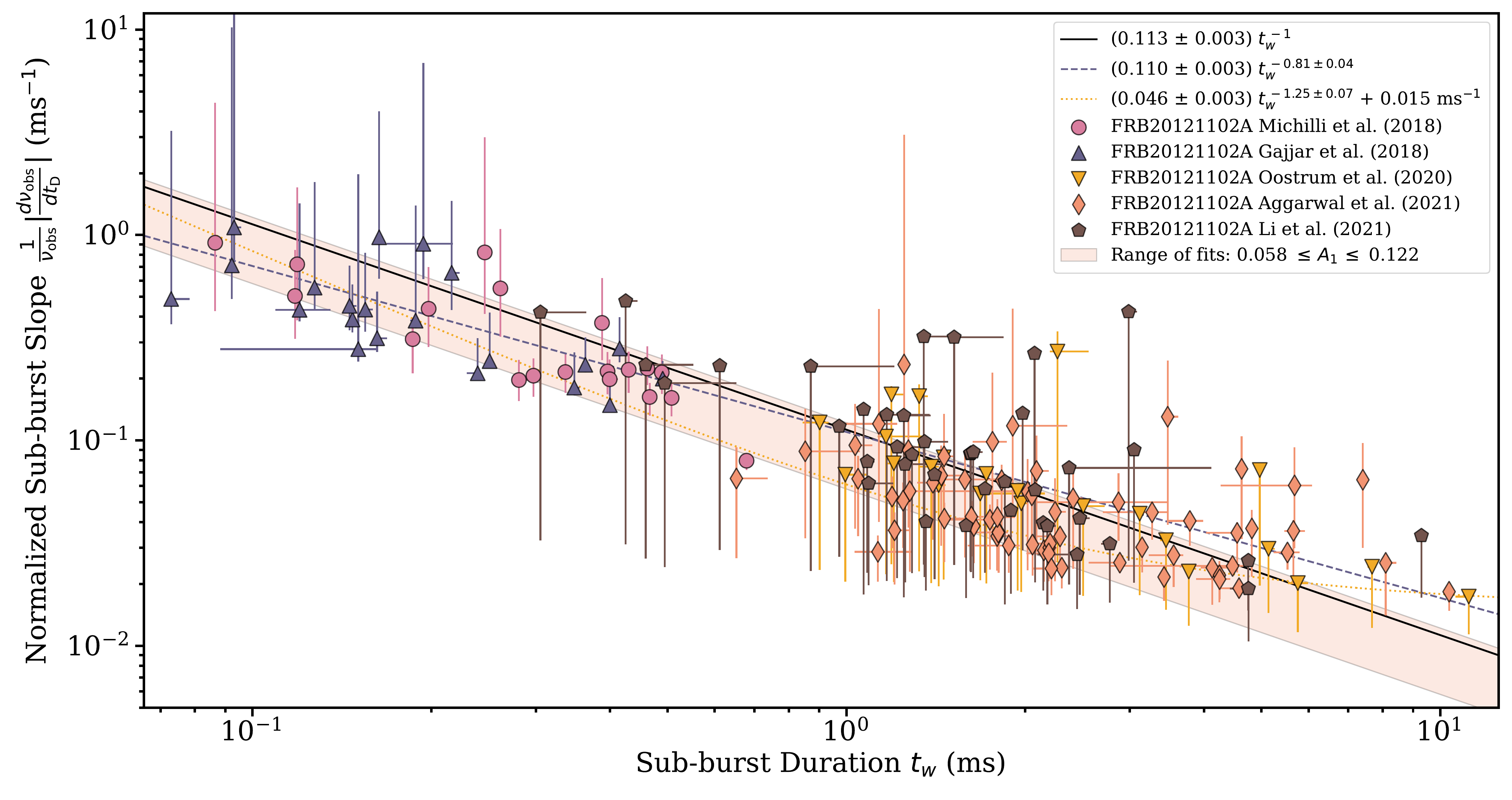}
    \caption{The relationship between the sub-burst slope and sub-burst duration. The sub-burst slope is normalized by the observed sub-burst frequency in order for the different datasets to be plotted together with the same relationship. The black line is a fit of the form $A\,t^{-1}_{\text{w}}$ which is the relationship predicted in the triggered relativistic dynamical model of \protect\cite{Rajabi2020}, while the other are more general fits of the form $A\,t^{n}_{\text{w}}$ and $A\,t^{n}_{\text{w}}+B$. The shaded region shows the intersection of the range of fits found over the DM ranges of each of the five datasets. Each measurement is displayed at the optimal DM for its source dataset, listed in Table \ref{tab:dmresults}. \review{Uncertainty bars show the range of measurements found over the limited DM range (see Section \ref{sec:methods}).} We see decent agreement between the three fits with the exponent of the expected relation lying between those found for the general fits.}
    \label{fig:SlopevDuration}
\end{figure*}

\begin{figure}
    \centering
    \includegraphics[width=\columnwidth]{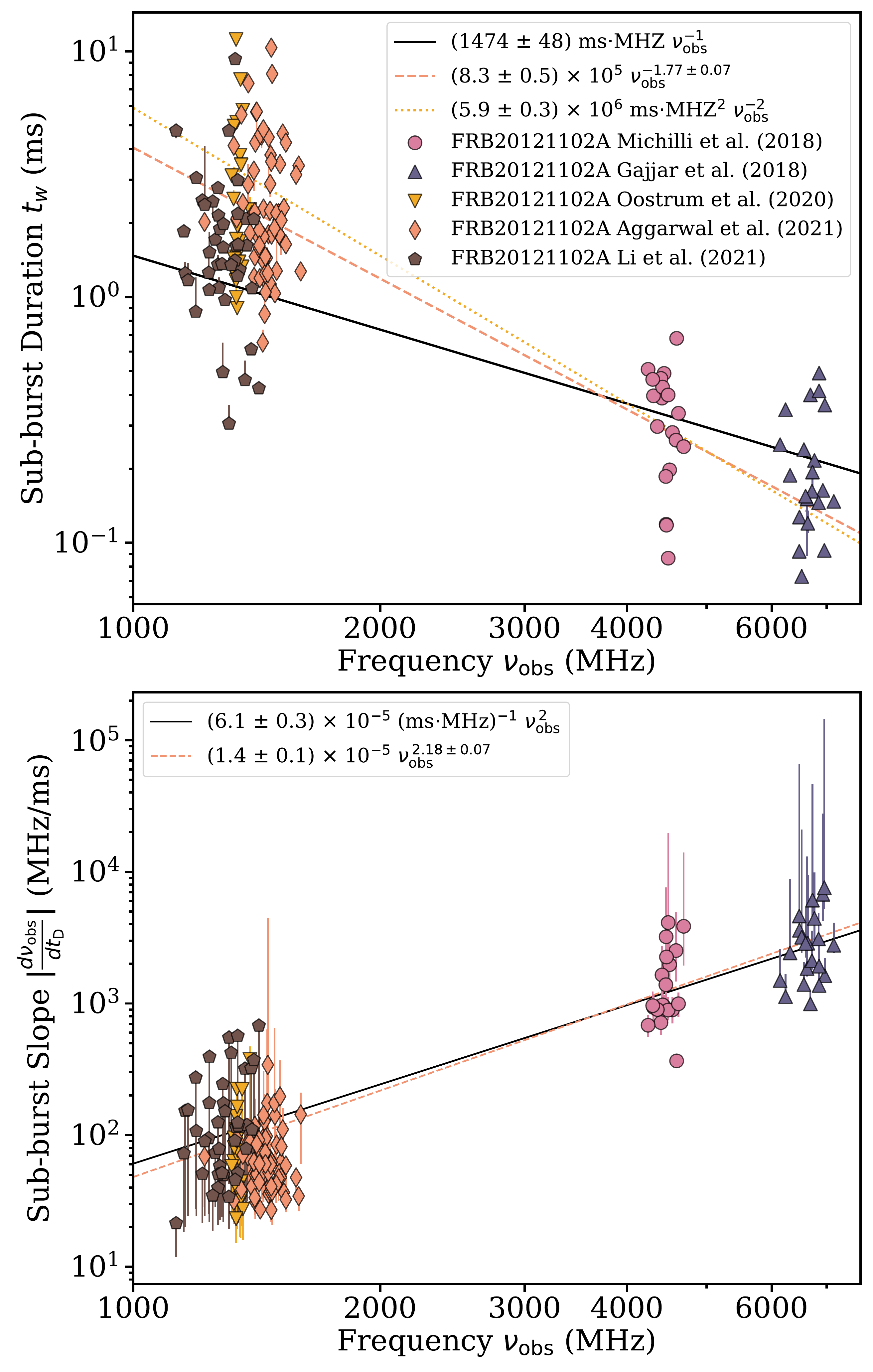}
    \caption{Relationships between the sub-burst duration (top) and the sub-burst slope (bottom) with burst frequency. Colors for all are shown in the top panel and are the same as in Figure \ref{fig:SlopevDuration}. In the top and bottom panels, the black line shows a fit to the relationship predicted by the triggered relativistic dynamical model in \protect\cite{Rajabi2020}, and the tan lines are general fits of the form $A\nu_{\text{obs}}^n$.}
    \label{fig:PropertiesvFreq}
\end{figure}

\begin{figure}
    \centering
    \includegraphics[width=\columnwidth]{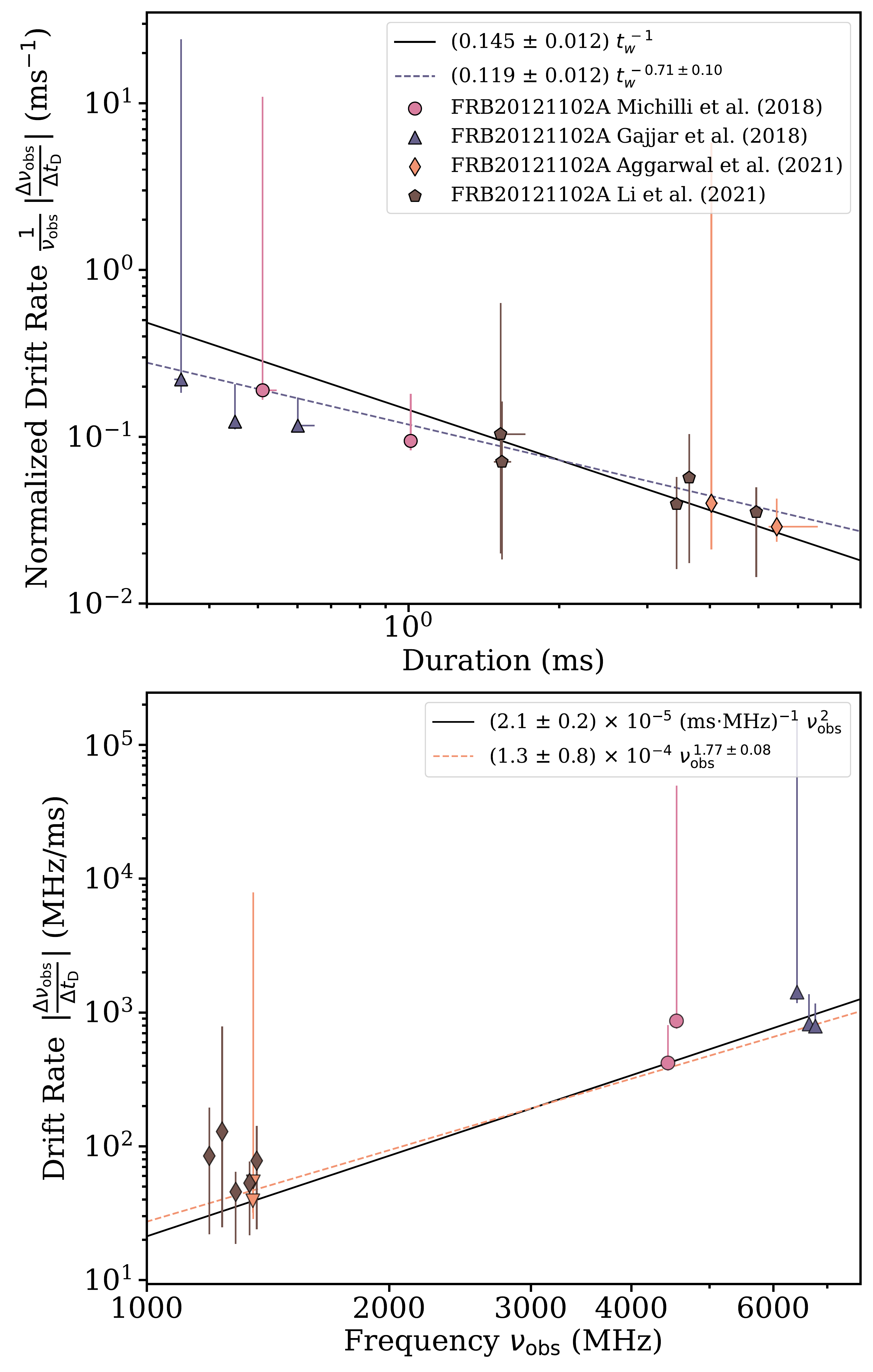}
    \caption{Relationships between the drift rate of multiple components and the duration (top) and the frequency of observation (bottom). On the left, fits of the form $A\,t^{-1}_{\text{w}}$ and $A\,t^{n}_{\text{w}}$ are made for comparison with the similar inverse relationship between the frequency-normalized sub-burst slope and duration shown in Figure \ref{fig:SlopevDuration}. On the right, we find two fits with one locked to a power of 2 and the other allowed to change.}
    \label{fig:Drifts}
\end{figure}

\begin{figure*}
    \centering
    \includegraphics[width=\textwidth]{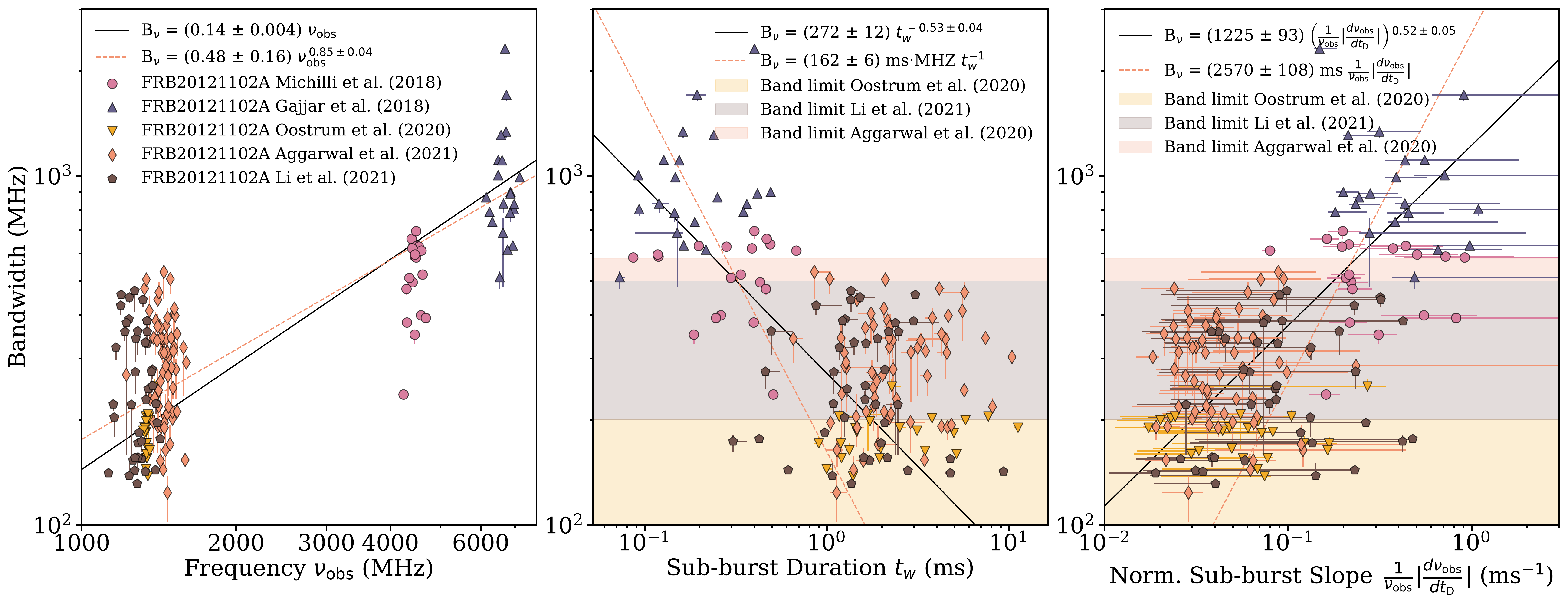}
    \caption{Relationships between the bandwidth and the burst frequency (left), sub-burst duration (middle) and sub-burst slope (right). The relationship between the bandwidth and frequency is linear, while the relationship with the duration goes, within uncertainties, to the power of $-$1/2. The origin of this relationship is unclear, and may be due to observational bias as indicated by the overlayed regions showing the bandwidth limits. \reviewb{The dashed lines in the middle and right panels show fits of the expected relationship implied by the left panel, and do not describe the data well.} In the right panel we see that the fit found between the bandwidth and slope is the inverse of that for the duration, as might be expected based on their inverse relationship shown in Figure \ref{fig:SlopevDuration}.}
    \label{fig:PropertiesvBand}
\end{figure*}

The correlations between the spectro-temporal properties shown in Figures \ref{fig:SlopevDuration} to \ref{fig:PropertiesvBand} confirm known relationships across a broad parameter range and reveal a potentially new relationship between the bandwidth and sub-burst duration (or normalized sub-burst slope). \review{A summary of the fits found is shown in Table \ref{tab:fitresults}}. A discussion of these results will be found in Section \ref{sec:discussion}.

\begin{table*}
\begin{centering}
\begin{tabular}{llrrrr}
\hline 
\textbf{\#} & \textbf{Form} & \textbf{$A$} & \textbf{$n$} & \textbf{$C$} & \textbf{Figure}\tabularnewline
\hline
\hline 
\textbf{1} & $\frac{1}{\nu_{\text{obs}}}\left|\frac{d\nu_{\text{obs}}}{dt_{\text{D}}}\right|=A_{1}t_{\text{w}}^{-1}$ & $0.113\pm0.003$ &  &  & \ref{fig:SlopevDuration}\tabularnewline
\hline 
\textbf{2} & $\frac{1}{\nu_{\text{obs}}}\left|\frac{d\nu_{\text{obs}}}{dt_{\text{D}}}\right|=A_{2}t_{\text{w}}^{n_{2}}$ & $0.110\pm0.003$ & $-0.81\pm0.04$ &  & \ref{fig:SlopevDuration}\tabularnewline
\hline 
\textbf{3} & $\frac{1}{\nu_{\text{obs}}}\left|\frac{d\nu_{\text{obs}}}{dt_{\text{D}}}\right|=A_{3}t_{\text{w}}^{n_{3}}+C_{3}$ & $0.046\pm0.003$ & $-1.25\pm0.07$ & $0.015\,\text{ms}^{-1}$ & \ref{fig:SlopevDuration}\tabularnewline
\hline 
\textbf{4} & $t_{\text{w}}=A_{4}\nu_{\text{obs}}^{-1}$ & $1474\pm48\,\text{ms\ensuremath{\cdot}MHz}$ &  &  & \ref{fig:PropertiesvFreq}\tabularnewline
\hline 
\textbf{5} & $t_{\text{w}}=A_{5}\nu_{\text{obs}}^{n_{5}}$ & $(8.3\pm0.5)\times10^5$ & $-1.77\pm0.07$ &  & \ref{fig:PropertiesvFreq}\tabularnewline
\hline 
\textbf{5b} & $t_{\text{w}}=A_{\text{5b}}\nu_{\text{obs}}^{-2}$ & $(5.9\pm0.3)\times10^{6}\,\text{ms\ensuremath{\cdot}MHz}^{2}$ &  &  & \ref{fig:PropertiesvFreq}\tabularnewline
\hline 
\textbf{6} & $\left|\frac{d\nu_{\text{obs}}}{dt_{\text{D}}}\right|=A_{6}\nu_{\text{obs}}^{2}$ & $(6.1\pm0.3)\times10^{-5}\,\text{ms}^{-1}\text{\ensuremath{\cdot}MHz}^{-1}$ &  &  & \ref{fig:PropertiesvFreq}\tabularnewline
\hline 
\textbf{7} & $\left|\frac{d\nu_{\text{obs}}}{dt_{\text{D}}}\right|=A_{7}\nu_{\text{obs}}^{n_{7}}$ & $(1.4\pm0.1)\times10^{-5}$ & $2.18\pm0.07$ &  & \ref{fig:PropertiesvFreq}\tabularnewline
\hline 
\textbf{8} & $\frac{1}{\nu_{\text{obs}}}\left|\frac{\Delta\nu_{\text{obs}}}{\Delta t_{\text{D}}}\right|=A_{8}t_{\text{w}}^{-1}$ & $0.145\pm0.012$ &  &  & \ref{fig:Drifts}\tabularnewline
\hline 
\textbf{9} & $\frac{1}{\nu_{\text{obs}}}\left|\frac{\Delta\nu_{\text{obs}}}{\Delta t_{\text{D}}}\right|=A_{9}t_{\text{w}}^{n_{9}}$ & $0.119\pm0.012$ & $-0.71\pm0.1$ &  & \ref{fig:Drifts}\tabularnewline
\hline 
\textbf{10} & $\left|\frac{\Delta\nu_{\text{obs}}}{\Delta t_{\text{D}}}\right|=A_{10}\nu_{\text{obs}}^{2}$ & $(2.1\pm0.2)\times10^{-5}\,\text{ms}^{-1}\text{\ensuremath{\cdot}MHz}^{-1}$ &  &  & \ref{fig:Drifts}\tabularnewline
\hline 
\textbf{11} & $\left|\frac{\Delta\nu_{\text{obs}}}{\Delta t_{\text{D}}}\right|=A_{11}\nu_{\text{obs}}^{n_{11}}$ & $(1.3\pm0.8)\times10^{-4}$ & $1.77\pm0.08$ &  & \ref{fig:Drifts}\tabularnewline
\hline 
\textbf{12} & $B_{\nu}=A_{12}\nu_{\text{obs}}$ & $0.14\pm0.004$ &  &  & \ref{fig:PropertiesvBand}\tabularnewline
\hline 
\textbf{13} & $B_{\nu}=A_{13}\nu_{\text{obs}}^{n_{13}}$ & $0.48\pm0.16$ & $0.85\pm0.04$ &  & \ref{fig:PropertiesvBand}\tabularnewline
\hline 
\textbf{14} & $B_{\nu}=A_{14}t_{\text{w}}^{n_{14}}$ & $272\pm12$ & $-0.53\pm0.04$ &  & \ref{fig:PropertiesvBand}\tabularnewline
\hline 
\textbf{15} & $B_{\nu}=A_{15}\left(\frac{1}{\nu_{\text{obs}}}\left|\frac{d\nu_{\text{obs}}}{dt_{\text{D}}}\right|\right)^{n_{15}}$ & $1225\pm93$ & $0.52\pm0.05$ &  & \ref{fig:PropertiesvBand}\tabularnewline
\hline 
\textbf{16} & $B_{\nu}=A_{16}t_{\text{w}}^{-1}$ & $162\pm6\,\text{ms\ensuremath{\cdot}MHz}$ &  &  & \ref{fig:PropertiesvBand}\tabularnewline
\hline 
\textbf{17} & $B_{\nu}=A_{17}\left(\frac{1}{\nu_{\text{obs}}}\left|\frac{d\nu_{\text{obs}}}{dt_{\text{D}}}\right|\right)$ & $2570\pm108\,\text{ms}$ &  &  & \ref{fig:PropertiesvBand}\tabularnewline
\hline
\end{tabular}
\par\end{centering}
\caption{A summary of fits attempted with the measurement results. The columns $A$, $n$, and $C$, respectively, show the amplitude of some kind of power-law, the
index of that power law, and any additive constants
when applicable. The variable $\nu_{\text{obs}}$ denotes the sub-burst frequency, $t_{\text{w}}$ denotes the sub-burst duration, and $(d\nu_{\text{obs}}/dt_{\text{D}})$ and $(\Delta\nu_{\text{obs}}/\Delta t_{\text{D}})$ denote the sub-burst slope and drift rate between multiple sub-bursts, respectively.}\label{tab:fitresults}
\end{table*}

In Figure \ref{fig:SlopevDuration} we plot the sub-burst slope normalized by the observing frequency versus the sub-burst duration. We find three fits to the data using the \texttt{scipy.odr} package, which performs an orthogonal distance regression, a method that takes into account the uncertainties on both variables. \review{We use the range of measurements over the DM range as the variable uncertainties when performing the fits.} The first fit is of the form $(d\nu_{\text{obs}}/dt_{\text{D}}) / \nu_{\text{obs}} = A_1\,t^{-1}_{\text{w}}$ which is the prediction of the TRDM and we find $A_1=0.113 \pm 0.003$. We fit two general power-laws, which are free of assumptions, of the form $A_2\,t^{n_2}_{\text{w}}$ and $A_3\,t^{n_3}_{\text{w}}+C_3$ to see \reviewc{how they compare} with the predicted fit. These yield $A_2=0.110 \pm 0.003,\, n_2=-0.81 \pm 0.04\,$ and $A_3=0.046 \pm 0.003,\, n_3=-1.25 \pm 0.07,\, C_3=0.015 \text{ms}^{-1}$. For each dataset, a range of fits of the form $A\,t^{-1}_{\text{w}}$ is found over the DM range and the shaded region shows the intersection of those ranges across the five datasets. That is, the shaded region shows the range of fits possible within the limited DM ranges of all the datasets and is an estimate of the uncertainty on the parameter $A_1$ as a function of the DM. We see that a majority of points fall within this region and are well described by the fits, with a small population of outliers located above the rest of the points in the sub-bursts with durations greater than 1 ms. \review{We also see the measurement ranges systematically tending to smaller sub-burst slopes with increasing duration.}

In Figure \ref{fig:PropertiesvFreq}, we plot the sub-burst duration and sub-burst slope against the frequency. In those two panels, the black line represents a fit of the form with the exponent fixed to the prediction of the TRDM and the tan line represents a fit where the exponent is free to vary. In the sub-burst duration vs. frequency plot, the black fit is of the form $A_4\,\nu_{\text{obs}}^{-1}$, while the tan fit leaves the index free and is of the form $A_5\,\nu_{\text{obs}}^{n_5}$. The free fit finds an exponent of $n_5=-1.77 \pm 0.07$, \reviewc{higher than} the predicted $n=-1$, with $A_4 = 1474 \pm 48\, \text{ms $\cdot$ MHz}$ and $A_5 = (8.3 \pm 0.5)\times10^5$. The yellow fit is of the form $A_{5\text{b}}\,\nu_{\text{obs}}^{-2}$, and will be discussed in Section \ref{subsec:bvtw}. In the sub-burst slope vs. frequency plot we fit to the TRDM form of $A_6\,\nu_{\text{obs}}^2$ and again to a free index form of $A_7\,\nu_{\text{obs}}^{n_7}$, finding $A_6 = (6.1 \pm 0.3)\times10^{-5} \text{ ms$^{-1}\cdot$ MHz$^{-1}$}$, $A_7 = (1.4 \pm 0.1) \times 10^{-5}$ and an index of $n_7=2.18 \pm 0.07$, \reviewc{slightly higher than} the predicted $n=2$ relationship. Both fits visually describe the data well, and we see a large spread in the bursts in both plots. 


The drift rate of a burst with multiple components may be due to a different phenomen\review{on} than that which results in the sub-burst slope, and so we separate the drift rate measurements from the sub-burst slope measurements and show their correlations in Figure \ref{fig:Drifts}. We find that their relationships are similar to those found for the sub-burst slopes. 

Of our sampled bursts only a handful showed multiple components. Most of these were measurable via the autocorrelation method but a few were not, such as in the cases where one of the bursts dominated the autocorrelation or when the drift rate was very small. The drift rates are put through the same filtering process as the sub-burst slopes and the top panel of Figure \ref{fig:Drifts} shows the normalized drift rate vs. the duration of the event, while the bottom \reviewc{panel} shows the drift rate vs. the frequency. As before, for the relationship with the duration we find one fit with the predicted form $A_{8}\,t_{\text{w}}^{-1}$ which yielded $A_{8}=0.145 \pm 0.012$ and a fit with a free exponent of the form $A_{9}\,t_{\text{w}}^{n_{9}}$ which yielded $A_{9}=0.119 \pm 0.012$ and $n_{9}=-0.71 \pm 0.1$. This first fit has an $A$ value higher than that found for the sub-burst slopes (\reviewc{i.e.} $A_{8} > A_1$) and outside the range of fits shown in Figure \ref{fig:SlopevDuration}. The second fit found an exponent $n_{9}$ with a smaller magnitude than $n_2$, the corresponding index for the sub-burst slopes. This suggests a relationship for the drift rates and duration similar to that found for the sub-burst slopes but with possibly a slightly different scaling. 

For the relationship between the drift rates and the frequency (bottom panel of Figure \ref{fig:Drifts}) we perform two fits, with the first fit of the TRDM form $A_{10}\,\nu_{\text{obs}}^2$ yielding $A_{10}=(2.1 \pm 0.2)\times10^{-5} \text{ ms$^{-1}\cdot$ MHz$^{-1}$}$, which is about a third the value found between the sub-burst slope and frequency ($A_6$, and shown in the bottom panel of Figure \ref{fig:PropertiesvFreq}). A fit of the form $A_{11}\,\nu_{\text{obs}}^{n_{11}}$ found $A_{11}=(1.3 \pm 0.8) \times 10^{-4}$ and $n_{11}=1.77 \pm 0.08$, with quite a large uncertainty on $A_{11}$. For the limited drift rates available, the fits to the drift rates' relationships are similar enough to those found for the sub-burst slopes that it is tempting to conclude that the same phenomena is responsible for both.

Figure \ref{fig:PropertiesvBand} shows plots of the sub-burst bandwidth against the frequency, sub-burst duration, and normalized sub-burst slope, respectively. For the bandwidth versus frequency plot we find a linear fit of the form $A_{12}\,\nu_{\text{obs}}$ consistent with Doppler broadening of narrowband emission and a free index fit of the form $A_{13}\,\nu_{\text{obs}}^{n_{13}}$. This yields $A_{12}=0.14 \pm 0.004$, $A_{13}=0.48 \pm 0.16$ and exponent $n_{13}=0.85 \pm 0.04$. For the fits to bandwidth versus the sub-burst duration and normalized slope (middle and right panels), we leave the exponent free and use the forms $A_{14}\,t_{\text{w}}^{n_{14}}$ and $A_{15}\,(\nu_{\text{obs}}^{-1}d\nu_{\text{obs}}/dt_{\text{D}})^{n_{15}}$. We find $A_{14}=272 \pm 12$ and $n_{14}=-0.53 \pm 0.04$ for the duration and $A_{15}=1225 \pm 93$ and $n_{15}=0.52 \pm 0.05$ for the normalized sub-burst slope. Plotting against the sub-burst slope instead of the normalized sub-burst slope has little effect on the exponent $n_{15}$. That the exponents $n_{14}$ and $n_{15}$ are inverses of each other within the uncertainties is expected from the inverse trend between the slope and duration shown in Figure \ref{fig:SlopevDuration}, in agreement with the prediction of the TRDM. 

That there is a relationship between the bandwidth and duration or slope of a burst can be expected from the other correlations shown but, as will be discussed in Section \ref{subsec:bvtw}, the exact form of relationship seen here is not, and is visible in these data because of the large range in frequency of the bursts sampled. However, as shown by the bandwidth limits plotted as regions in the figures, the maximum observable bandwidths for the \cite{Oostrum2020}, \cite{Aggarwal2021}, and \cite{Li2021} datasets are 200 MHz, \reviewc{580} MHz, and 500 MHz, respectively, due to their instrumentation (with a few exceptions for bursts that had slightly different backends), implying that sub-bursts with high bandwidths and long durations could be missed. \reviewc{This however does not explain the observed absence of low bandwidth, short duration bursts.}

The nearest and simplest fractional value consistent with the uncertainties of the exponent $n_{14}=-0.53 \pm 0.04$ is $-$1/2 and so the two rightmost panels of Figure \ref{fig:PropertiesvBand}, if observationally complete, suggest a proportionality of $B_{\nu} \propto t^{-1/2}_{\text{w}} \propto (\nu_{\text{obs}}^{-1}d\nu_{\text{obs}}/dt_{\text{D}})^{1/2}$, where $B_{\nu}$ is the bandwidth. The colored lines in the two rightmost panels show fits of the form $A_{16}\,t^{-1}_{\text{w}}$ and $A_{17}\,(\nu_{\text{obs}}^{-1}d\nu_{\text{obs}}/dt_{\text{D}})$, respectively, with $A_{16}=162 \pm 6$ ms$\cdot$MHz and $A_{17} = 2570 \pm 108$ ms. These \reviewc{last two fits} are the expected relationships inferred from our results \reviewc{and will} be discussed in Section \ref{subsec:bvtw} where we also further explore this unusual dependence. 

The set of correlations described above provide evidence that the relationships predicted by the TRDM between the sub-burst duration, slope and frequency apply for a majority of bursts from this source and possibly indicate an additional unknown relationship between the bandwidth and sub-burst slope/duration.

\begin{figure*}
    \centering
    \includegraphics[width=\textwidth]{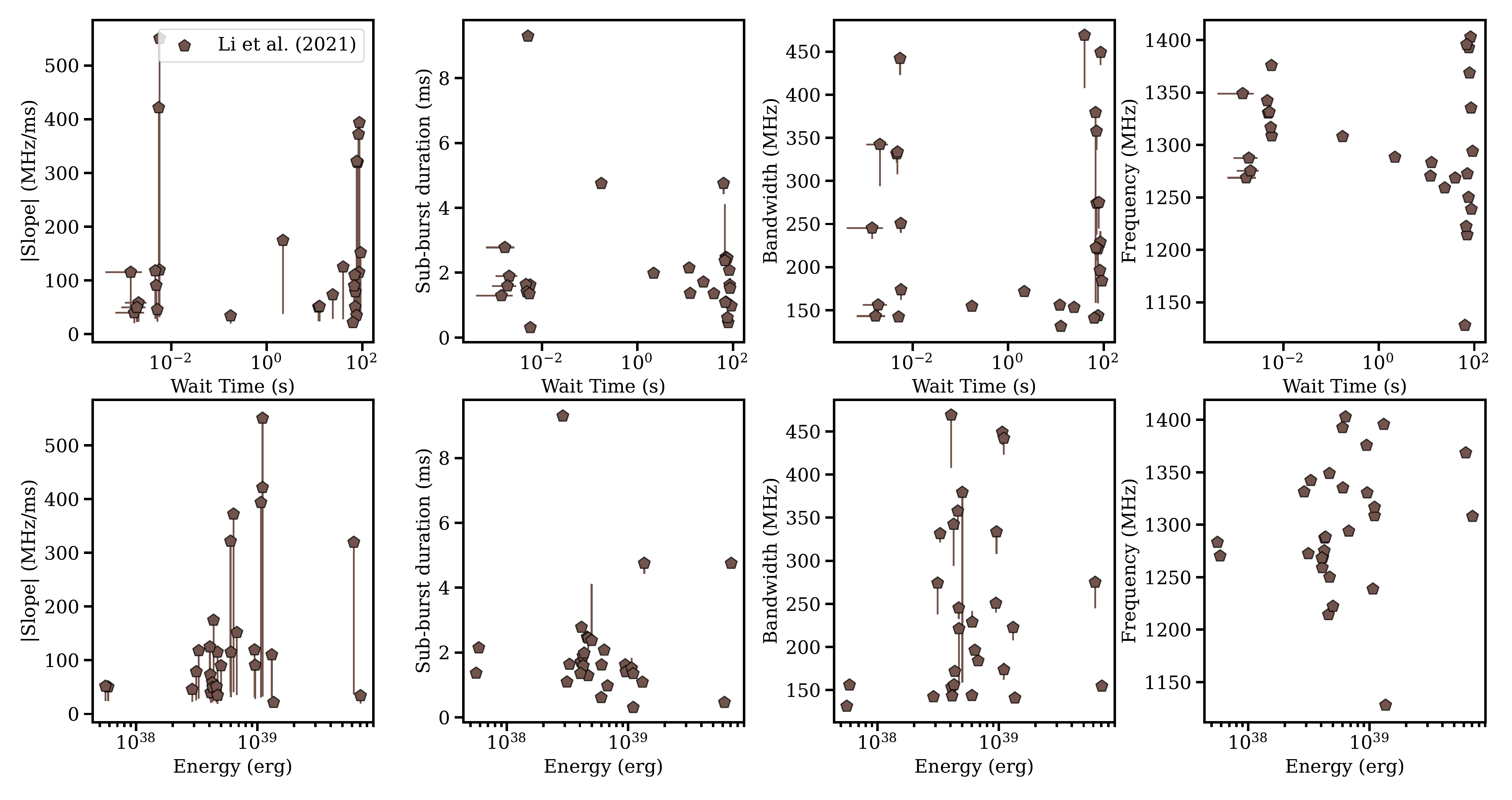}
    \caption{Search for correlations between measured spectro-temporal properties of 28 bursts from \protect\cite{Li2021} and their wait-times (top) and energies (bottom), as reported by those authors. \review{This figure does not include the 13 sub-bursts that were separated ourselves.} The two peaks of the wait-time distributions can be seen in the top row around 10$^{-2}$ and 10$^2$ seconds. The two peaks of the energy distribution are less obvious due to difficulties measuring bursts from the low energy ($<10^{38}$ erg) peak \review{and the small sample limits the possible conclusions}. No obvious \review{dependence} of a burst's spectro-temporal properties on either its wait-time or energy is seen in these data.}
    \label{fig:Li}
\end{figure*}

For the sampled bursts from the bimodal peaks in energy and wait-time distributions reported by \cite{Li2021}, we searched for correlations between our measured spectro-temporal properties and the burst energy and wait-time, finding that they were unrelated to each other. Figure \ref{fig:Li} shows plots of the slope, sub-burst duration, bandwidth, and frequency as functions of the wait-time (top row) and burst energy (bottom row). In the top row we can see that the bursts cluster around wait-times of $10^{-2}$ and $10^{2}$ seconds, corresponding to the two sampled peaks, and that the spectro-temporal measurements in all panels vary randomly and with similar ranges. In the bottom row a large number of bursts cluster around $10^{39}$ erg while the low energy peak at $<10^{38}$ erg only has two points, due to difficulties we encountered in measuring these lower S/N bursts. Nonetheless, across the four panels we again see that the measurements vary widely at the high energy peak and cover similar values as the measurements from the low energy peak. For this limited sample of data we therefore conclude that no clear relationship exists between the wait-time and energy of a burst with any of its spectro-temporal properties. 

\section{Discussion}\label{sec:discussion}
We discuss here our results in the context of other studies and the TRDM, including the expected relationships for the drift rates and their connection to those for the sub-burst slopes. We will also discuss the unexpected relationship observed between the bandwidth, duration, and normalized sub-burst slope, the \reviewc{possible} implications of the relationships observed (and not observed, in the case of the spectro-temporal properties with the burst energy and wait-time), as well as the large spread in values seen in the measurements of the spectro-temporal properties and what these results mean for understanding FRBs overall.


Several studies have now closely examined the relationships between spectro-temporal properties of FRBs and found strong correlations between them, especially between the normalized sub-burst slope and sub-burst duration. The first report of the inverse relationship between the slope and duration was in \cite{Rajabi2020}, seen with 25 bursts from FRB~20121102A. \cite{Chamma2021} reported the same relationship for multiple sources seen with bursts from FRB20180916B and FRB20180814A \reviewc{as well as} FRB~20121102A, while also reporting that the same \reviewc{value of the} fit parameter $A$ characterized the relationship for the three sources. In earlier work the data for FRB~20121102A were limited to bursts with short durations and frequencies greater than 4 GHz, and since then large numbers of bursts were made available at longer durations and lower frequencies, presenting the opportunity to understand the relationships seen earlier with improved statistics. 

In this study the frequency and duration ranges are broad and the number of bursts studied from FRB~20121102A is large enough to conclude that a majority of bursts from this source follow a tight relationship between the normalized sub-burst slope and duration. The optimum fit parameter found here was $A_1=0.113 \pm 0.003$, which is higher than the $A=0.078 \pm 0.006$ found by \cite{Chamma2021}, along with a narrower range of fits centered around $A=0.09$. \reviewb{We note} the range of fits includes this earlier value. 

In \cite{Jahns2022}, 849 bursts from FRB~20121102A were analysed on the lower end of our frequency range (around 1400 MHz) and a strong inverse correlation between the sub-burst duration and slope can be seen in their Figure 5. Their formalism finds the inverse of the sub-burst slope relation used here and their fit parameter of $|b|=0.00862(37)$ MHz$^{-1}$ can be converted to \reviewc{our formalism} for comparison via $A=1/(\nu_{\text{obs}}b)$, which yields $A_{\text{Jahns}}=0.0839$ when using $\nu_{\text{obs}}=1400$ MHz, the most common burst frequency in that sample. This value is comparable to the results found here, and differences can be due to the burst definition used and details of the analysis (such as normalizing each point by the burst frequency). 

The TRDM predicts the relationship between the sub-burst duration and frequency to be
\begin{align}
    t_{\text{w}}=\tau'_{\text{w}}\frac{\nu_{0}}{\nu_{\text{obs}}},\label{eq:twmodel}
\end{align}
where $t_{\text{w}}$ is the sub-burst duration, $\tau'_{\text{w}}$ is the sub-burst duration in the rest frame of emission, $\nu_0$ is the rest frame frequency, and $\nu_{\text{obs}}$ is the frequency in the observer frame. The values of $\tau'_{\text{w}}$ and $\nu_0$ can be absorbed into a constant and so an inverse relationship is expected. Other evidence of this relationship can be found in Figure 7 of \cite{Gajjar2018} and Figure 3 of \cite{Hessels2019}. Each `clump' of bursts seen in Figure \ref{fig:PropertiesvFreq} viewed independently appears uncorrelated with the observing frequency, but placed together a downward trend becomes clear. 

The relationship between sub-burst slope and frequency is similar in this regard as well, since over a narrow spectral range the two observables can appear uncorrelated. \cite{Wang2022} in their Figure 8 using data from multiple studies and multiple FRB sources and spanning a large frequency range found that trends proportional to $\nu^2$ and $\nu^{2.29}$ fit the data well. A quadratic relationship between sub-burst slope and frequency was previously implied by the results of \cite{Chamma2021} in their Figure 5. Our results here \reviewc{again show that a quadratic} relationship fit\reviewc{s} well, \reviewc{with} the free exponent fit \reviewc{finding} a value of $\nu^{2.18 \pm 0.07}$. 

The relationship between the sub-burst slope and frequency was derived in \cite{Chamma2021} from the TRDM using equation (\ref{eq:twmodel}) and the relationship between the sub-burst slope and duration, \reviewc{which} is given by
\begin{align}
    \frac{1}{\nu_{\text{obs}}}\frac{d\nu_{\text{obs}}}{dt_{\text{D}}}=-\left(\frac{\tau'_{\text{w}}}{\tau'_{\text{D}}}\right)\frac{1}{t_{\text{w}}}=-\frac{A}{t_{\text{w}}},\label{eq:slopemodel}
\end{align}
where as before $d\nu_{\text{obs}}/dt_{\text{D}}$ is the sub-burst slope, $t_\text{D}$ is the delay time (or arrival time), and $\tau'_{\text{D}}$ is the delay time in the rest frame of the FRB. The fit parameter $A$ is seen here as the ratio between the rest frame sub-burst duration and delay time, and this relationship is the one predicted for the sub-burst slope and duration, shown with the black fit in Figure \ref{fig:SlopevDuration}. Combining equations (\ref{eq:twmodel}) and (\ref{eq:slopemodel}) we obtain 
\begin{align}
    \frac{d\nu_{\text{obs}}}{dt_{\text{D}}}=-\frac{A\nu_{\text{obs}}^{2}}{\tau'_{\text{w}}\nu_{0}}\equiv-A_6\,\nu_{\text{obs}}^{2},\label{eq:slopevfreq}
\end{align}
where the new constant $A_6=A/(\tau'_{\text{w}}\nu_0)=1/(\tau'_\text{D}\nu_0)$. This is a quadratic relationship between the sub-burst slope and frequency and is \reviewc{consistent with} our results as well as the results of \cite{Chamma2021} and \cite{Wang2022}. 

The TRDM assumes a fundamentally narrow-band emission process at a rest frame frequency $\nu_0$ and if the rest frame sub-burst duration $\tau'_\text{w}$ can be estimated with a more detailed model of the emission process, then the ratio of $A$ and $A_6$ would provide a measure of $\nu_0$, through 
\begin{align}
    \frac{A}{A_6}=\nu_{0}\tau'_{\text{w}}.\label{eq:nu0ratio}
\end{align}
For example using our results of $A_1=0.113 \pm 0.003$ and $A_6 = (6.1 \pm 0.3)\times10^{-5} \text{ ms$^{-1}\cdot$ MHz$^{-1}$}$ and with an order of magnitude estimate of $\tau'_{\text{w}}\approx 1 \,\text{ms}$, we can very roughly estimate $\nu_0$ to be on the order 1 GHz. 

The TRDM offers a simple explanation for the relationships \review{discussed here through its assumption of} a narrow-band emission process for FRBs at a common rest frame frequency $\nu_0$.

\subsection{Drift rate vs duration and frequency}
\review{As discussed in Section \ref{sec:results},} the drift rate relationships with duration and frequency are plotted in Figure \ref{fig:Drifts} and these relationships appear to be analogous to those for the sub-burst slope. For the limited data available, the drift rate relationship with frequency seems to be adequately described by the fit proportional to $\nu_{\text{obs}}^2$, as the free fit is proportional to $\nu_{\text{obs}}^{1.77}$, which is closer to quadratic than not. We can derive a relationship between the drift rate and the frequency using equation 8 of \cite{Rajabi2020}, which describes the change in frequency with arrival time of distinct sub-bursts. Expanding the $d\nu_{\text{obs}}/d\tau'_{\text{D}}$ term in that equation we get
\begin{align}
    \frac{\Delta\nu_{\text{obs}}}{\Delta t_{\text{D}}}=\frac{\nu_{\text{obs}}}{\nu_{0}}\frac{d\nu_{\text{obs}}}{d\tau'_{\text{D}}}=\frac{\nu_{\text{obs}}}{\nu_{0}}\frac{d\nu_{\text{obs}}}{d\beta}\frac{d\beta}{d\tau'_{\text{D}}},\label{eq:driftdbeta}
\end{align}
where $\Delta \nu_\text{obs} / \Delta t_\text{D}$ is the drift rate and $\beta$ is the fraction of the speed of light the FRB source is moving with along the line of sight. By using the relativistic Doppler shift formula $\nu_{\text{obs}}=\nu_0\sqrt{(1+\beta)/(1-\beta)}$, we can evaluate $d\nu_{\text{obs}}/d\beta$ in the above expression to obtain
\begin{align}
    \frac{\Delta\nu_{\text{obs}}}{\Delta t_{\text{D}}}=\frac{\nu_{\text{obs}}^{2}}{\nu_{0}}\gamma^{2}\frac{d\beta}{d\tau'_{\text{D}}},\label{eq:driftvfreq}
\end{align}
where $\gamma=1/\sqrt{1-\beta^2}$. Equation (\ref{eq:driftvfreq}) shows a $\nu_{\text{obs}}^2$ dependence with the frequency with $\gamma^{2}d\beta/d\tau'_{\text{D}}$ a parameter characterizing the region under study. More data are required to \review{verify what relationship exists between the drift rate and frequency, and the result of such an analysis} can be interpreted within the model based on the relations described above. The data also seem to suggest an inverse relationship analogous to the relationship between the sub-burst slope and sub-burst duration. As mentioned earlier, this can be expected in the TRDM when groups of sub-bursts are emitted at roughly the same time \citep[Sec. 3.1 of][]{Chamma2021,Rajabi2020}.

\subsection{Bandwidth vs. Sub-burst Duration relationship}\label{subsec:bvtw}
The correlations of spectro-temporal properties with the bandwidth shown in Figure \ref{fig:PropertiesvBand} reveal known as well as unknown relationships. One known spectro-temporal correlation seen in FRBs is the scaling of bandwidth with frequency. For example, in \cite{Houde2019}, the linear relationship shown in the \reviewb{left} panel of Figure \ref{fig:PropertiesvBand} between bandwidth and frequency is seen in their Figure 5, and includes points between 2 and 4 GHz. Their slope, once converted for comparison is $0.156$, close to the value $A_{12} = 0.14 \pm 0.004$ obtained here. This linear relationship between the bandwidth and frequency follows from the non- or weakly-relativistic Doppler effect \citep{Houde2019}. 

The relationships found in the \reviewb{middle and right} panels of Figure \ref{fig:PropertiesvBand} where $B_{\nu} \propto t^{-1/2}_{\text{w}}$ and $B_{\nu} \propto (d\nu_{\text{obs}}/dt_{\text{D}})^{1/2}$ are \reviewc{inconsistent with the TRDM}, difficult to verify, and statistically significant enough to be considered real. Based on our results \reviewc{as well as the predictions of the TRDM}, the expected dependence is $B_{\nu} \propto t^{-1}_{\text{w}}$ \reviewc{consistent with} the dependencies $t_{\text{w}} \propto \nu_{\text{obs}}^{-1}$ (top panel of Figure \ref{fig:PropertiesvFreq}) and $B_{\nu} \propto \nu_{\text{obs}}$ (\reviewc{left} panel of Figure \ref{fig:PropertiesvBand}). Accordingly, the expected dependence between $B_{\nu}$ and $d\nu_{\text{obs}}/dt_{\text{D}}$ is linear. However, fits of the form $B_{\nu} \propto t^{-1}_{\text{w}}$ and $B_{\nu} \propto d\nu_{\text{obs}}/dt_{\text{D}}$, shown by the \reviewb{dashed} lines in the \reviewb{middle and right} panels of Figure \ref{fig:PropertiesvBand}, fail to simultaneously cross both the low frequency and high frequency bursts as well as the $B_{\nu} \propto t^{-1/2}_{\text{w}}$ and $B_{\nu} \propto (d\nu_{\text{obs}}/dt_{\text{D}})^{1/2}$ fits do. 

\reviewc{In an effort to explain this relationship outside of the TRDM, we can take advantage of the fact that} the bandwidth appears to be linear with frequency \reviewc{and} perform a test fit on the sub-burst duration and frequency that assumes the form $t_{\text{w}} \propto \nu_{\text{obs}}^{-2}$ since this would imply a bandwidth-duration relationship of the form $B_{\nu} \propto t^{-1/2}_{\text{w}}$ consistent with our \reviewc{observations}. Indeed, the $\nu_{\text{obs}}^{-2}$ fit (shown in the first panel of Figure \ref{fig:PropertiesvFreq}) is adequate enough of a fit that we cannot be entirely certain whether $t_{\text{w}} \propto \nu^{-1}_{\text{obs}}$ or $t_{\text{w}} \propto \nu^{-2}_{\text{obs}}$. However, we note that the relationship $t_{\text{w}} \propto \nu^{-1}_{\text{obs}}$ is at the basis for the sub-burst slope law, i.e., $(d\nu_{\text{obs}}/dt_{\text{D}}) / \nu_{\text{obs}} \propto t^{-1}_{\text{w}}$, through a simple derivative, and that this relation appears to be securely confirmed here and elsewhere \citep{Rajabi2020,Chamma2021,Jahns2022}. \reviewc{On the other hand, we note that} a combination of the sub-burst slope law with $t_{\text{w}} \propto \nu^{-2}_{\text{obs}}$ leads to a relationship $d\nu_{\text{obs}}/dt_{\text{D}} \propto \nu^{3}_{\text{obs}}$, \reviewc{that is not quite} compatible with \reviewc{our observed results pointing to a quadratic relationship}. \reviewc{If} we are inclined to favour a $t_{\text{w}} \propto \nu^{-1}_{\text{obs}}$ relation \reviewc{then this can naturally explain the sub-burst slope law}. \reviewc{In that case,} the $B_{\nu} \propto t^{-1/2}_{\text{w}}$ functionality \reviewc{appears at odds with the TRDM. If instead we favour the $t_\text{w} \propto \nu^{-2}_\text{obs}$ relation, then we lack an explanation for the sub-burst slope law and we are in tension with the observed evidence for a quadratic relationship between the slope and frequency}.

There is no known relationship within the TRDM that explains the $B_{\nu} \propto t^{-1/2}_{\text{w}}$ relationship between the bandwidth and the sub-burst duration, or the $B_{\nu} \propto \left[(d\nu_{\text{obs}}/dt_{\text{D}})/\nu_{\text{obs}}\right]^{1/2}$ relationship between the bandwidth and sub-burst slope. If the velocity distribution of the FRB source, which determines the bandwidth in the TRDM, is an independent variable that helps in determining the sub-burst duration and sub-burst slope, then it is not clear what physical mechanism enables the velocity of one sample of material to determine the duration of the pulse emitted by material moving at a different velocity. 


T\reviewb{he \reviewc{absence} of a readily available explanation for the bandwidth-duration/slope relationship from our results suggests that the explanation may be related to the physics of the emission process. \reviewc{However, a resolution to this inconsistency could involve changing or discarding the TRDM, ruling out the observed bandwidth-duration relationship, or some mixture of both.}}


\subsection{Spectro-temporal properties and their relationship to the emission process}
Across the different relationships explored and the broad sample of bursts considered from FRB~20121102A, there does not seem to be a large population of outliers that breaks sharply from the relationships described\review{. This fact,} in light of the \review{absence} of correlations between the spectro-temporal properties and the energies and wait-times, suggests \review{that} some of the spectro-temporal properties of an FRB and its emission mechanism \review{\reviewc{could be} independent of one another}. \reviewc{If we were to assume the predictions of the TRDM are consistently true, then} a simple \review{interpretation} for the \reviewc{absence} of outliers is that the correlations between properties like the sub-burst slope, duration and frequency of FRBs (at least from this source) originate from the relativistic dynamics of the source as explained by the TRDM. \reviewc{Under this assumption,} the correlation between the bandwidth and duration, unexplained in the TRDM as discussed above, falls into a different category. \reviewc{Of course, another interpretation is that the TRDM is incomplete.}

\reviewc{However, }\review{t}he results presented here \reviewc{are evidence} that the FRB emission from this source comes from material that is moving relativistically, since several spectro-temporal correlations of FRB~20121102A are explainable through the dynamics of that material \review{and the TRDM}, and not due to the actual emission process. 

This \review{view} is supported by the results shown in Figure \ref{fig:Li}, where several spectro-temporal properties appear uncorrelated with both the wait-time and energy of a burst. Further supporting examples are the plots with energy displayed in Figure 5 of \cite{Jahns2022} for bursts around 1400 MHz, the relationship between the frequency and flux density shown in Figure 4 of \cite{Gajjar2018} for bursts around 7 GHz, and plots of different properties with the fluence shown in Figure 5 of \cite{Aggarwal2021} for bursts around 1400 MHz, all of which imply no clear relationship between the spectro-temporal properties and burst energies. 

These aforementioned studies are done independently over a narrow range of frequencies and durations. An analysis with a broad sample such as the one used for this study that focuses on the energy and/or fluxes of bursts would be needed to confirm the \reviewc{absence} of correlation between burst energies and spectro-temporal properties such as the sub-burst slope or duration, as the present results and literature suggest. \reviewc{If so, and assuming the validity of the TRDM, then such an absence would be evidence that what we observe from FRBs can be broken into two regimes; properties which are determined from the motion of the source (TRDM), and properties which are determined from the so far unknown emission process.}

\reviewc{We also note }there is considerable statistical spread in the values of sub-burst slope, duration and bandwidth seen in Figures \ref{fig:SlopevDuration} to \ref{fig:PropertiesvBand} \review{from each dataset} that could be related to the emission process. For example, the data for \cite{Michilli2018} cluster around 4.5 GHz, but values for the sub-burst slope range from $5\times10^2$ MHz/ms to $4\times10^3$ MHz/ms, and the sub-burst duration spans nearly an order of magnitude. Similar statements are true for all the other datasets considered. In the TRDM the sub-burst duration and sub-burst slope (see equations \ref{eq:twmodel} and \ref{eq:slopemodel}) are related to the rest frame frequency $\nu_0$, and the rest frame duration and delay times $\tau'_{\text{w}}$ and $\tau'_{\text{D}}$. The dependence of the sub-burst duration and slope on these properties, which in the TRDM are determined exclusively by the emission process and FRB environment, is possibly the cause of the \review{scatter} seen in the spectro-temporal properties. Characterizing the \review{scatter} of spectro-temporal properties for large samples of bursts may therefore be an \reviewc{additional} avenue for studying the emission process.








\section{Conclusions}\label{sec:conclusions}
We \reviewc{have} \review{measured} the spectro-temporal properties and the relationships between them of a broad sample of 167 bursts from FRB~20121102A with frequencies ranging from 1--7.5 GHz and sub-burst durations ranging from less than 1 ms to about 10 ms from the observational studies of \cite{Michilli2018}, \cite{Gajjar2018}, \cite{Oostrum2020}, \cite{Aggarwal2021}, and \cite{Li2021}. \review{Bursts from the latter study were sampled} from the two peaks of the bimodal energy and bimodal wait-time distributions reported therein in order to search for relationships between those parameters and the spectro-temporal properties of the bursts. Our measurements of the spectro-temporal properties include measurements of the sub-burst slope, sub-burst duration, burst bandwidth, center frequency, and sad trombone drift rate (when applicable) at each DM between 555 to 575 pc/cm$^{3}$ in steps of 0.5 pc/cm$^{3}$. The complete set of measurements are available online along with an extensible graphical user interface called \textsc{Frbgui} that was developed and used to prepare and perform measurements on the burst waterfalls. 

We characterized multiple relationships between the measured spectro-temporal properties of bursts from FRB~20121102A finding general agreement with multiple predictions from the TRDM described in \cite{Rajabi2020}, which fundamentally assumes a narrow-band emission process and up-to relativistic motion. We found, as in \cite{Rajabi2020}, \cite{Chamma2021} and \cite{Jahns2022}, an inverse relationship between the sub-burst slope and sub-burst duration. In addition, we find an inverse relation between the sub-burst duration and frequency, a quadratic relation between the sub-burst slope and frequency, and a linear relationship between the sub-burst bandwidth and frequency.

The 12 drift rates measured were seen to follow relationships with the duration and frequency that are analogous to the ones obeyed by the sub-burst slope. However, more drift rates are needed to properly characterize these trends. 

We also found an unexpected correlation with power-law index consistent with $-$1/2 between the sub-burst bandwidth and sub-burst duration $(B_{\nu} \propto t_{\text{w}}^{-1/2})$. \reviewc{Consistent with} the inverse relationship between the sub-burst slope and duration, \reviewc{we also found} a relationship between the bandwidth and sub-burst slope with power-law index 1/2 (i.e. $B_{\nu} \propto \left[(d\nu_{\text{obs}}/dt_{\text{D}})/\nu_{\text{obs}}\right]^{1/2}$). This relationship is \reviewc{hard to explain within the TRDM} and is potentially uncertain due to the limited observing bands at lower frequencies, however this would not explain the \reviewc{absence} of \reviewc{low} bandwidth, short duration sub-bursts observed at high frequencies. 

\review{We }found \review{no correlations between} the spectro-temporal properties of a burst and its energy or wait-time.

Except for the bandwidth relationship with the normalized sub-burst slope and duration, the multiple relationships observed between the spectro-temporal properties generally agree with the predictions made in \cite{Rajabi2020}. Therefore, more data are needed from FRB~20121102A between 2--4 GHz and high bandwidth bursts should be searched for at low frequencies (below 2 GHz), as well as low bandwidth bursts at high frequencies, in order to test the limits of the bandwidth-duration/slope relationship found. \reviewc{Such observations are also needed to pin down the relationship between the sub-burst duration and frequency, and sub-burst slope and frequency.}

The general agreement of the spectro-temporal properties with the predictions of the TRDM from this broad sample of bursts from FRB~20121102A suggests that several of the spectro-temporal relationships of all bursts from FRB~20121102A can be explained by dynamical motions of the FRB source, \reviewc{though there still remain challenges to validating all the predictions of the TRDM. No significant population of outliers was found with our burst sample and the correlations studied were sufficiently characterized by a single relationship.} Meanwhile, the \reviewc{scatter} of values observed in the spectro-temporal properties \reviewc{could} arise from parameters in the model that are determined exclusively by the emission mechanism such as the rest frame frequency of emission and the rest frame timescales. Therefore, characterizing the \reviewc{scatter} of spectro-temporal properties from large samples of bursts may also inform on the emission mechanism of FRBs. 



\section*{Acknowledgements}
We are grateful to Di Li for arranging access to and support with the FAST data. The authors are grateful to Christopher Wyenberg for feedback and discussions on drafts of this manuscript \reviewb{and to the referee for their \reviewc{careful reading} and suggestions on improving clarity.} M.H.'s research is funded through the Natural Sciences and Engineering Research Council of Canada Discovery Grant RGPIN-2016-04460. Plots in the paper were prepared using the Matplotlib package \citep{Hunter2007}.

\section*{Data Availability}
 
All spectro-temporal measurements are available online at \href{https://github.com/mef51/SurveyFRB20121102A}{\texttt{github.com/mef51/SurveyFRB20121102A}}. The waterfalls of the FRBs can be obtained from the authors of their respective publications. The graphical user interface \textsc{Frbgui} and related scripts used to perform the measurements are available at \href{https://github.com/mef51/frbgui}{\texttt{github.com/mef51/frbgui}}. 



\bibliographystyle{mnras}
\bibliography{main} 

\begin{thebibliography}{}
\makeatletter
\relax
\def\mn@urlcharsother{\let\do\@makeother \do\$\do\&\do\#\do\^\do\_\do\%\do\~}
\def\mn@doi{\begingroup\mn@urlcharsother \@ifnextchar [ {\mn@doi@}
  {\mn@doi@[]}}
\def\mn@doi@[#1]#2{\def\@tempa{#1}\ifx\@tempa\@empty \href
  {http://dx.doi.org/#2} {doi:#2}\else \href {http://dx.doi.org/#2} {#1}\fi
  \endgroup}
\def\mn@eprint#1#2{\mn@eprint@#1:#2::\@nil}
\def\mn@eprint@arXiv#1{\href {http://arxiv.org/abs/#1} {{\tt arXiv:#1}}}
\def\mn@eprint@dblp#1{\href {http://dblp.uni-trier.de/rec/bibtex/#1.xml}
  {dblp:#1}}
\def\mn@eprint@#1:#2:#3:#4\@nil{\def\@tempa {#1}\def\@tempb {#2}\def\@tempc
  {#3}\ifx \@tempc \@empty \let \@tempc \@tempb \let \@tempb \@tempa \fi \ifx
  \@tempb \@empty \def\@tempb {arXiv}\fi \@ifundefined
  {mn@eprint@\@tempb}{\@tempb:\@tempc}{\expandafter \expandafter \csname
  mn@eprint@\@tempb\endcsname \expandafter{\@tempc}}}

\bibitem[\protect\citeauthoryear{Agarwal et~al.,}{Agarwal
  et~al.}{2020}]{Agarwal2020}
Agarwal D.,  et~al., 2020, \mn@doi [Monthly Notices of the Royal Astronomical
  Society] {10.1093/mnras/staa1927}, 497, 352

\bibitem[\protect\citeauthoryear{Aggarwal}{Aggarwal}{2021}]{Aggarwal2021b}
Aggarwal K.,  2021, \mn@doi [The Astrophysical Journal Letters]
  {10.3847/2041-8213/ac2a3a}, 920, L18

\bibitem[\protect\citeauthoryear{Aggarwal, Agarwal, Lewis, Anna-Thomas,
  Tremblay, Burke-Spolaor, McLaughlin  \& Lorimer}{Aggarwal
  et~al.}{2021}]{Aggarwal2021}
Aggarwal K.,  Agarwal D.,  Lewis E.~F.,  Anna-Thomas R.,  Tremblay J.~C.,
  Burke-Spolaor S.,  McLaughlin M.~A.,   Lorimer D.~R.,  2021, \mn@doi [The
  Astrophysical Journal] {10.3847/1538-4357/ac2577}, 922, 115

\bibitem[\protect\citeauthoryear{Anna-Thomas et~al.,}{Anna-Thomas
  et~al.}{2022}]{AnnaThomas2022}
Anna-Thomas R.,  et~al., 2022, arXiv

\bibitem[\protect\citeauthoryear{Bassa et~al.,}{Bassa et~al.}{2017}]{Bassa2017}
Bassa C.~G.,  et~al., 2017, \mn@doi [The Astrophysical Journal]
  {10.3847/2041-8213/aa7a0c}, 843, L8

\bibitem[\protect\citeauthoryear{CHIME/FRB}{CHIME/FRB}{2020}]{CHIME2020b}
CHIME/FRB 2020, \mn@doi [Nature] {10.1038/s41586-020-2398-2}, 582, 351

\bibitem[\protect\citeauthoryear{{{CHIME/FRB Collaboration}}
  et~al.,}{{{CHIME/FRB Collaboration}} et~al.}{2019}]{Amiri2019}
{{CHIME/FRB Collaboration}} et~al., 2019, \mn@doi [Nature]
  {10.1038/s41586-018-0864-x}, 566, 235

\bibitem[\protect\citeauthoryear{{CHIME/FRB Collaboration} et~al.,}{{CHIME/FRB
  Collaboration} et~al.}{2021}]{CHIMEFRB2021}
{CHIME/FRB Collaboration} et~al., 2021, \mn@doi [\apjs]
  {10.3847/1538-4365/ac33ab}, 257, 59

\bibitem[\protect\citeauthoryear{Chamma, Rajabi, Wyenberg, Mathews  \&
  Houde}{Chamma et~al.}{2021}]{Chamma2021}
Chamma M.~A.,  Rajabi F.,  Wyenberg C.~M.,  Mathews A.,   Houde M.,  2021,
  \mn@doi [Monthly Notices of the Royal Astronomical Society]
  {10.1093/mnras/stab2070}, 507, 246

\bibitem[\protect\citeauthoryear{Chatterjee et~al.,}{Chatterjee
  et~al.}{2017}]{Chatterjee2017}
Chatterjee S.,  et~al., 2017, \mn@doi [\nat] {10.1038/nature20797}, 541, 58

\bibitem[\protect\citeauthoryear{Gajjar et~al.,}{Gajjar
  et~al.}{2018}]{Gajjar2018}
Gajjar V.,  et~al., 2018, \mn@doi [The Astrophysical Journal]
  {10.3847/1538-4357/aad005}, 863, 2

\bibitem[\protect\citeauthoryear{Gourdji, Michilli, Spitler, Hessels, Seymour,
  Cordes  \& Chatterjee}{Gourdji et~al.}{2019}]{Gourdji2019}
Gourdji K.,  Michilli D.,  Spitler L.~G.,  Hessels J. W.~T.,  Seymour A.,
  Cordes J.~M.,   Chatterjee S.,  2019, \mn@doi [The Astrophysical Journal]
  {10.3847/2041-8213/ab1f8a}, 877, L19

\bibitem[\protect\citeauthoryear{Hessels et~al.,}{Hessels
  et~al.}{2019}]{Hessels2019}
Hessels J. W.~T.,  et~al., 2019, \mn@doi [The Astrophysical Journal]
  {10.3847/2041-8213/ab13ae}, 876, L23

\bibitem[\protect\citeauthoryear{Hilmarsson et~al.,}{Hilmarsson
  et~al.}{2021}]{Hilmarsson2021}
Hilmarsson G.~H.,  et~al., 2021, \mn@doi [The Astrophysical Journal]
  {10.3847/2041-8213/abdec0}, 908, L10

\bibitem[\protect\citeauthoryear{Houde, Rajabi, Gaensler, Mathews  \&
  Tranchant}{Houde et~al.}{2019}]{Houde2019}
Houde M.,  Rajabi F.,  Gaensler B.~M.,  Mathews A.,   Tranchant V.,  2019,
  \mn@doi [Monthly Notices of the Royal Astronomical Society]
  {10.1093/mnras/sty3046}, 482, 5492

\bibitem[\protect\citeauthoryear{Hunter}{Hunter}{2007}]{Hunter2007}
Hunter J.~D.,  2007, \mn@doi [Computing in Science and Engineering]
  {10.1109/MCSE.2007.55}, 9, 90

\bibitem[\protect\citeauthoryear{Jahns et~al.,}{Jahns et~al.}{2022}]{Jahns2022}
Jahns J.~N.,  et~al., 2022, arXiv

\bibitem[\protect\citeauthoryear{Josephy et~al.,}{Josephy
  et~al.}{2019}]{Josephy2019}
Josephy A.,  et~al., 2019, \mn@doi [The Astrophysical Journal]
  {10.3847/2041-8213/ab2c00}, 882, L18

\bibitem[\protect\citeauthoryear{Li et~al.,}{Li et~al.}{2021}]{Li2021}
Li D.,  et~al., 2021, \mn@doi [Nature] {10.1038/s41586-021-03878-5}, 598, 267

\bibitem[\protect\citeauthoryear{Luo et~al.,}{Luo et~al.}{2020}]{Luo2020}
Luo R.,  et~al., 2020, \mn@doi [Nature] {10.1038/s41586-020-2827-2}, 586, 693

\bibitem[\protect\citeauthoryear{Lyubarsky}{Lyubarsky}{2021}]{Lyubarsky2021}
Lyubarsky Y.,  2021, \mn@doi [Universe] {10.3390/universe7030056}, 7, 56

\bibitem[\protect\citeauthoryear{Marcote et~al.,}{Marcote
  et~al.}{2017}]{Marcote2017}
Marcote B.,  et~al., 2017, \mn@doi [\apjl] {10.3847/2041-8213/834/2/L8}, 834,
  L8

\bibitem[\protect\citeauthoryear{Michilli et~al.,}{Michilli
  et~al.}{2018}]{Michilli2018}
Michilli D.,  et~al., 2018, \mn@doi [Nature] {10.1038/nature25149}, 553, 182

\bibitem[\protect\citeauthoryear{Nimmo et~al.,}{Nimmo et~al.}{2022}]{Nimmo2022}
Nimmo K.,  et~al., 2022, \mn@doi [Nature Astronomy]
  {10.1038/s41550-021-01569-9}, 6, 393

\bibitem[\protect\citeauthoryear{Nita, Gary  \& Hellbourg}{Nita
  et~al.}{2016}]{Nita2016}
Nita G.~M.,  Gary D.~E.,   Hellbourg G.,  2016, in 2016 Radio Frequency
  Interference ({RFI}). {IEEE}, \mn@doi{10.1109/rfint.2016.7833535}

\bibitem[\protect\citeauthoryear{Oostrum et~al.,}{Oostrum
  et~al.}{2020}]{Oostrum2020}
Oostrum L.~C.,  et~al., 2020, \mn@doi [Astronomy {\&} Astrophysics]
  {10.1051/0004-6361/201937422}, 635, A61

\bibitem[\protect\citeauthoryear{Petroff, Hessels  \& Lorimer}{Petroff
  et~al.}{2022}]{Petroff2022}
Petroff E.,  Hessels J. W.~T.,   Lorimer D.~R.,  2022, \mn@doi [The Astronomy
  and Astrophysics Review] {10.1007/s00159-022-00139-w}, 30

\bibitem[\protect\citeauthoryear{Platts, Weltman, Walters, Tendulkar, Gordin
  \& Kandhai}{Platts et~al.}{2019}]{Platts2019}
Platts E.,  Weltman A.,  Walters A.,  Tendulkar S.,  Gordin J.,   Kandhai S.,
  2019, \mn@doi [Physics Reports] {10.1016/j.physrep.2019.06.003}, 821, 1

\bibitem[\protect\citeauthoryear{Pleunis et~al.,}{Pleunis
  et~al.}{2021}]{Pleunis2021a}
Pleunis Z.,  et~al., 2021, \mn@doi [The Astrophysical Journal]
  {10.3847/1538-4357/ac33ac}, 923, 1

\bibitem[\protect\citeauthoryear{Rajabi, Chamma, Wyenberg, Mathews  \&
  Houde}{Rajabi et~al.}{2020}]{Rajabi2020}
Rajabi F.,  Chamma M.~A.,  Wyenberg C.~M.,  Mathews A.,   Houde M.,  2020,
  \mn@doi [Monthly Notices of the Royal Astronomical Society]
  {10.1093/mnras/staa2723}, 498, 4936

\bibitem[\protect\citeauthoryear{Spitler et~al.,}{Spitler
  et~al.}{2014}]{Spitler2014}
Spitler L.~G.,  et~al., 2014, \mn@doi [The Astrophysical Journal]
  {10.1088/0004-637x/790/2/101}, 790, 101

\bibitem[\protect\citeauthoryear{Spitler et~al.,}{Spitler
  et~al.}{2016}]{Spitler2016}
Spitler L.~G.,  et~al., 2016, \mn@doi [Nature] {10.1038/nature17168}, 531, 202

\bibitem[\protect\citeauthoryear{Tendulkar et~al.,}{Tendulkar
  et~al.}{2017}]{Tendulkar2017}
Tendulkar S.~P.,  et~al., 2017, \mn@doi [\apjl] {10.3847/2041-8213/834/2/L7},
  834, L7

\bibitem[\protect\citeauthoryear{Wang, Yang, Niu, Xu  \& Zhang}{Wang
  et~al.}{2022}]{Wang2022}
Wang W.-Y.,  Yang Y.-P.,  Niu C.-H.,  Xu R.,   Zhang B.,  2022, \mn@doi [The
  Astrophysical Journal] {10.3847/1538-4357/ac4097}, 927, 105

\makeatother
\end{thebibliography}








\bsp	
\label{lastpage}
\end{document}